\documentclass[usenatbib,useAMS,fleqn]{mnras}

\usepackage{newtxtext,newtxmath}

\usepackage[T1]{fontenc}
\usepackage{ae,aecompl}

\usepackage{tikz, graphicx}
\usepackage{amsmath}	
\usepackage{amssymb}	
\usepackage{multicol}        
\usepackage{threeparttable}
\usepackage{xspace}
\usepackage{times}

\usepackage{etoolbox}
\makeatletter
\patchcmd\@combinedblfloats{\box\@outputbox}{\unvbox\@outputbox}{}{\errmessage{\noexpand\@combinedblfloats could not be patched}}
\makeatother

\newcommand{\spit}{{\it Spitzer}} 
\newcommand{\her}{{\it Herschel}} 
\newcommand{\sofia}{{\it SOFIA}} 

\newcommand{\go}{$G_{\rm{UV}}$}

\newcommand{\gstar}{$G_{\rm stars}$}
\newcommand{\ff}{$\Phi_{\rm A}$}                                 
\newcommand{\Av}{$A_{V}$}                                      
\newcommand{\Avmax}{$A_{V}^{\rm max}$}            
\newcommand{\Lfir}{$L_{\rm FIR}$}                                  
\newcommand{\fir}{L_{\rm FIR}}                               
\newcommand{\Ltir}{$L_{\rm TIR}$}                                  
   
\newcommand{\mic}{\,$\mu$m}                                          
\newcommand{\cm}{\,cm$^{-3}$}                                        
\newcommand{\wmsqsr}{${\rm W m}^{-2}{\rm sr}^{-1}$} 
\newcommand{\kms}{\,km\,s$^{-1}$}  
\newcommand{\Zsun}{{\,\rm Z}_\odot} 
\newcommand{ \Xco}{$X_{\rm CO}$}   
\newcommand{ \Xcomw}{$X_{\rm CO, MW}$}   
\newcommand{ \Xcodor}{$X_{\rm CO, 30Dor}$}   
\newcommand{ \Aco}{$\alpha_{\rm CO}$}   
\newcommand{ \Acomw}{$\alpha_{\rm CO, MW}$}   
\newcommand{\Xunit}{\,cm$^{-2}$ (K km s$^{-1}$)$^{-1}$}       
\newcommand{\Aunit}{\,M$_\odot$ pc$^{-2}$ (K km s$^{-1}$)$^{-1}$}       

\newcommand{\Cp}{C$^+$}            
\newcommand{\Co}{C$^{\rm 0}$}    
\newcommand{\HH}{H$_2$}          
\newcommand{\HI}{H$\,${\sc i}}     
\newcommand{\HII}{H$\,${\sc ii}}     
\newcommand{\OIII}{[O$\,${\sc iii}]}    
\newcommand{\SII}{[S$\,${\sc ii}]}    
\newcommand{\SIII}{[S$\,${\sc iii}]}    
\newcommand{\OI}{[O$\,${\sc i}]} 
\newcommand{\CII}{[C$\,${\sc ii}]} 
\newcommand{\CI}{[C$\,${\sc i}]} 
\newcommand{\NII}{[N$\,${\sc ii}]} 

\title[The CO-dark molecular gas mass in 30 Doradus]{The CO-dark molecular gas mass in 30 Doradus}

\author[M.~Chevance et al.]{M\'elanie Chevance,$^{1}$\thanks{chevance@uni-heidelberg.de}
Suzanne C.~Madden,$^{2}$
Christian Fischer,$^{3}$
William D.~Vacca,$^{4}$
\newauthor
Vianney Lebouteiller,$^{2}$
Dario Fadda,$^{4}$
Fr\'ed\'eric Galliano,$^{2}$
Remy Indebetouw,$^{5,6}$
\newauthor
J.~M.~Diederik Kruijssen,$^{1}$
Min-Young Lee,$^{7}$
Albrecht Poglitsch,$^{8}$
Fiorella L.~Polles,$^{9}$
\newauthor
Diane Cormier,$^{2}$
Sacha Hony,$^{10}$
Christof Iserlohe,$^{3}$
Alfred Krabbe,$^{3}$
Margaret Meixner,$^{11}$
\newauthor
Elena Sabbi,$^{11}$
Hans Zinnecker$^{12}$
\vspace{2mm}
\\
$^{1}$Astronomisches Rechen-Institut, Zentrum f\"ur Astronomie der Universit\"at Heidelberg, M\"onchhofstrasse 12-14, 69120 Heidelberg, Germany.\\
$^{2}$AIM, CEA, CNRS, Universit\'e Paris-Saclay, Universit\'e Paris Diderot, Sorbonne Paris Cit\'e, 91191, Gif-sur-Yvette, France\\
$^{3}$Deutsches SOFIA Institut, Pfaffenwaldring 29, D-70569 Stuttgart, Germany\\
$^{4}$SOFIA-USRA, NASA Ames Research Center, MS N232-12, Moffett Field, CA 94035-1000, USA\\
$^{5}$Department of Astronomy, University of Virginia, Charlottesville, VA 22904, USA\\ 
$^{6}$National Radio Astronomy Observatory, 520 Edgemont Road, Charlottesville, VA 22903, USA\\
$^{7}$Korea Astronomy and Space Science Institute, 776 Daedeokdae-ro, 34055, Daejeon, Republic of Korea\\
$^{8}$Max-Planck-Institut f\"ur extraterrestrische Physik, Gie\ss enbachstrasse 1, D-85748 Garching, Germany\\
$^{9}$LERMA, Observatoire de Paris, PSL Research University, CNRS, Sorbonne Universit\'e, 75014, Paris, France\\
$^{10}$Institut f\"ur theoretische Astrophysik, Zentrum f\"ur Astronomie der Universit\"at Heidelberg, Albert-Ueberle Str. 2, 69120, Heidelberg, Germany\\
$^{11}$Space Telescope Science Institute, Baltimore, MD 21218, USA\\
$^{12}$Universidad Autonoma de Chile Av Pedro de Valdivia 425, Providencia, Santiago de Chile, Chile
}
\date{Accepted 2020 April 16. Received 2020 April 13; in original form 2020 March 23.}

\pubyear{2020}

\begin{document}
\label{firstpage}
\pagerange{\pageref{firstpage}--\pageref{lastpage}}
\maketitle

\begin{abstract}
Determining the efficiency with which gas is converted into stars in galaxies requires an accurate determination of the total reservoir of molecular gas mass. However, despite being the most abundant molecule in the Universe, \HH\ is challenging to detect through direct observations and indirect methods have to be used to estimate the total molecular gas reservoir. These are often based on scaling relations from tracers such as CO or dust, and are generally calibrated in the Milky Way. Yet, evidence that these scaling relations are environmentally dependent is growing. In particular, the commonly used CO-to-\HH\ conversion factor (\Xco ) is expected to be higher in metal-poor and/or strongly UV-irradiated environments. We use new \sofia/FIFI-LS observations of far-infrared fine structure lines from the ionised and neutral gas and the Meudon photodissociation region model to constrain the physical properties and the structure of the gas in the massive star-forming region of 30 Doradus in the Large Magellanic Cloud, and determine the spatially resolved distribution of the total reservoir of molecular gas in the proximity of the young massive cluster R136. We compare this value with the molecular gas mass inferred from ground-based CO observations and dust-based estimates to quantify the impact of this extreme environment on commonly used tracers of the molecular gas. We find that the strong radiation field combined with the half-solar metallicity of the surrounding gas are responsible for a large reservoir of "CO-dark" molecular gas, leaving a large fraction of the total \HH\ gas ($\gtrsim 75$\%) undetected when adopting a standard \Xco\ factor in this massive star-forming region.
\end{abstract}



\section{Introduction}
\label{sec:intro}

The strong correlation between tracers of star formation and tracers of molecular gas observed on kpc scales in galaxies \citep[e.g.][]{Kennicutt1998, Wong2002, Leroy2008, Bigiel2011} suggests that star formation is mostly fueled by molecular gas \citep[e.g.][]{Krumholz2005, Elmegreen2007, McKee2007, Krumholz2009, ostriker2010}. While \cite{Glover2012b} show that this correlation does not necessarily imply a direct causation, the density and shielding conditions required for star formation are similar to those for the formation of molecular hydrogen, \HH\ \citep[e.g.][]{Krumholz2011, Glover2012b, Glover2012}. This is the most abundant molecule in the interstellar medium (ISM), and is generally a major constituent of the star-forming gas. However, \HH\ is a symmetric molecule, with no electric dipole moment. The lowest rotation-vibrational transition is thus the quadrupole transition at $\sim$28\mic\ (h$\nu$/k $\sim$ 500\,K), which is not excited in the cold gas phase associated with star formation ($\sim$ 10\,K). Other molecules, such as carbon monoxide (CO), are easily excited even at low temperatures, and are used as a proxy for the total reservoir of molecular gas, both in local galaxies \citep[e.g.][]{ Bigiel2011, Leroy2011, Schruba2012, Saintonge2017}, and at high redshift \citep[e.g.][]{Daddi2010, Walter2014, Genzel2015, Pavesi2018, Tacconi2018}. The emission of the CO $J=1-0$ transition at 2.6 mm, hereafter CO(1-0), is frequently used as a tracer of the molecular gas mass. A simple CO-to-\HH\ conversion factor \Xco\ is used to infer the \HH\ column density, $N(\text{\HH})$, from the CO(1-0) line intensity, $I_{\text{CO(1-0)}}$, such that $N(\text{\HH}) = \text{\Xco} \times I_{\text{CO(1-0)}}$. In the Milky Way, the common adopted value for this conversion factor is \Xcomw ~$= 2 \times 10^{20}$ \Xunit\  \citep[see e.g][and references therein]{Bolatto2013}.

While CO seems to trace the bulk of \HH\ in solar metallicity galaxies relatively well \citep[e.g.][]{Bolatto2013}, detection of CO in dwarf galaxies is challenging, even for metallicities only a few times below solar \citep[e.g.][]{Madden2013, Cormier2014}. In particular, CO emission has been shown to be disproportionately weak in the Large Magellanic Cloud (LMC) and other low-metallicity galaxies \citep{Wong2011, Hughes2010} compared to their star formation activities (estimated via \CII\ 158\mic\ emission or far-infrared luminosity; e.g. \citealt{Madden2013, Cormier2014, DeLooze2014, Cormier2015}). By contrast, the star formation activity of some low-metallicity galaxies as probed by the ratio of the [CII] luminosity over the CO luminosity \citep[e.g.][]{Stacey1991} can be very high \citep{Madden2000}, as is particularly the case for blue compact dwarf galaxies. As a result, star-forming low-metallicity galaxies are often outliers on the Schmidt-Kennicutt relation \citep{Kennicutt1989}, displaying lower gas surface densities at a given star formation rate density than solar metallicity galaxies (e.g. \citealt{Taylor1998, Galametz2009, Schruba2012, Cormier2014}). Whether this results from an decreased gas depletion time (and in this case, \HH\ is indeed less abundant than expected from solar metallicity galaxy relations) or from the fact that CO fails at being an effective tracer of the total molecular gas in these environments (with the \HH\ reservoir and the gas depletion time actually being comparable to normal galaxies) remains unanswered.

Indeed, the \Xco\ factor is expected to vary with environmental properties, and notorious uncertainties exist when inferring the \HH\ reservoir from CO, especially in low-metallicity environments. While a constant \Xco\ factor, or a linear scaling of this value with metallicity, is often assumed to derive the total mass of molecular gas, these values remain rough estimates  (e.g.\ \citealt{Schruba2012}, see also the review by \citealt{Bolatto2013}). These uncertainties strongly limit our interpretation of the evolution of star formation activity and gas mass at high redshift \citep[e.g.][]{Daddi2010, Tacconi2010, Genzel2012}. Indeed, contrarily to \HH , which is efficiently self-shielded, CO is photodissociated by the radiation field at low dust column densities. Therefore, \Xco\ has been suggested to increase in metal-poor and/or highly irradiated environments, where the transition between C$^+$/C$^0$/CO is shifted deeper into the cloud, whereas a substantial reservoir of \HH , called "CO-dark" molecular gas, can be found outside the CO-emitting regions (where carbon is found in atomic or ionised form; e.g.~\citealt{Israel1997, Wolfire2010, Bolatto2013}, Madden et al.~in prep.).

Other indirect probes of the molecular gas have been used to trace this potentially important reservoir of \HH . These include $\gamma$-ray emission \citep[e.g][]{Grenier2005}, dust emission {with an assumed dust-to-gas mass ratio} \citep[e.g][]{Planck2011}, or emission from the \CI\ line  \citep[e.g][]{Papadopoulos2004, Bell2006, Bell2007, Valentino2018}. However, the calibration of these diagnostics as a function of the environment also remains uncertain. For example, the dust-to-gas mass ratio has also been shown to depend strongly on the metallicity \citep{Roman-Duval2014,Remy-Ruyer2014,Remy-Ruyer2015} and shows significant scatter between galaxies, {most likely due to different star formation histories \citep[see][for a review]{Galliano2018}.}

The question remains as to whether there is a true deficit of \HH\ in low-metallicity environments relative to their star formation activity, or a lack an accurate measurement of the gas reservoir under lower dust shielding conditions. Low-metallicity galaxies are often characterised by higher and harder radiation fields (e.g. \citealt{Madden2006, Cormier2015, Cormier2019}), and a reduced dust content (e.g. \citealt{Remy-Ruyer2014}), which both contribute to changing the structure of the ISM, making it in particular more porous to photons than in higher metallicity galaxies \citep{Cormier2015, Chevance2016}. These properties affect the star formation process {by potentially reducing the reservoir of gas available for star formation}, and influence how the stellar radiation interacts with the surrounding medium. These characteristics may resemble the ones of the chemically young medium \citep[$\sim 0.4\Zsun$, e.g.][]{Yuan2013, Ly2016} found in galaxies at redshift $\sim 2-3$. This makes nearby low-metallicity star-forming dwarf galaxies perfect laboratories to investigate these questions at high spatial resolution. This is a critical step in order to obtain a better understanding of the conditions under which most of the stars in the Universe formed (at $z \sim 2-3$, \citealt{Madau2014}). Eventually, nearby dwarf galaxies can be used to calibrate new diagnostics for the total reservoir of molecular gas as a function of the environment \citep[e.g.][]{Lupi2020}.

In this paper we focus on the extreme environment around the Super Stellar Cluster (SSC) R136 on the 30 Doradus region (hereafter "30Dor") in the LMC. At a distance of 50\,kpc \citep{Walker2012}, the LMC is our closest neighbouring galaxy, and has a moderately low metallicity of $0.5 \Zsun$ \citep{Rolleston2002, Pagel2003}. 30Dor is the most prominent and most massive star-forming region in the LMC. The SSC R136 in its centre contains a large population of massive stars \citep[see the recent review by][and references therein]{Crowther2019}, irradiating the surrounding gas. In a previous paper (\citealt{Chevance2016}), we have combined atomic fine-structure line observations of 30Dor from \her\ and \spit\ with ground-based CO data to provide diagnostics on the properties and the structure of the gas, by modeling it with the ''Meudon'' model of photodissociation regions (PDR)\footnote{The Meudon PDR model is public and available online at the following address: \url{http://ism.obspm.fr.}} \citep{LePetit2006, Bron2014}. In the study by \cite{Lee2019}, we expanded the neutral gas tracers to include additional CO lines accessible from the \her /SPIRE FTS observations of the 30Dor region. 

The observed (or modeled) reservoir of molecular gas in a single region is dependent on the current evolutionary stage of the region \citep[e.g.][]{Kruijssen2018}. A small molecular gas mass can indeed result from \HH\ having been consumed by the on-going star formation event, or having been photodissociated by the stellar feedback from young massive stars, while a large \HH\ reservoir potentially indicates that the star formation event just started in this region. Measuring the total molecular gas mass in a single region therefore does not allow us to determine whether the star formation is more efficient at low metallicity. However, this is a first step towards a better understanding of how stellar feedback interacts with the surrounding sub-solar metallicity gas.

In the current paper, we determine additional constraints on the physical conditions of the ionised gas with a larger spatial coverage achieved by recent observations with the Stratospheric Observatory for Infrared Astronomy (\sofia ; \citealt{Young2012, Krabbe2013}) of FIR fine structure lines from the ionised and neutral gas. We make use of these new observations to predict the distribution of the total \HH\ column density and mass surface density in this region. We present the data in Section~\ref{sec:observations} and discuss empirical diagnostics of the ISM conditions in Section~\ref{sec:lines}. In Section~\ref{sec:PDR}, we use the Meudon PDR model to calculate the mass surface density and total mass of \HH\ in 30Dor and determine the \Xco\ factor for this region, by comparison to the standard Galactic value \Xcomw . We discuss environmental variations of this measurement and compare it with literature values in Section~\ref{sec:discussion}. Key results and conclusions are summarised in Section~\ref{sec:concl}.

\section{Observations}
\label{sec:observations}

We present here the observational data used in this study to constrain the physical conditions of the gas in 30Dor and predict the total reservoir of molecular gas. We use new observations of FIR fine structure lines from the ionised and neutral gas with \sofia , in addition to archival spectrometric and photometric observations from \her\ and molecular gas observations from Mopra.
 
\subsection{SOFIA/FIFI-LS}
\label{sec:obs_FIFI}

We present new observations of the \OIII\ lines at 52\mic\ and 88\mic, the \OI\ line at 145\mic\ and the \CII\ line at 158\mic , taken with the Far Infrared Field-Imaging Line Spectrometer (FIFI-LS) instrument \citep{Krabbe2013, Fischer2018, Colditz2018} aboard \sofia . FIFI-LS is an imaging spectrometer with a wavelength dependent spectral resolution $R$ between $\sim$ 500 and $\sim$ 2000. The instrument provides two channels for simultaneous observations between 51-120\mic\ (blue channel) and 115-203\mic\ (red channel). Each channel has a $5 \times 5$ pixel projection onto the sky, with a size of 6\arcsec$\times$6\arcsec\ ($\sim$1.5\,pc $\times$ 1.5\,pc) in the blue channel and 12\arcsec$ \times$12\arcsec\ ($\sim$2.9\,pc $\times$ 2.9\,pc) in the red channel, yielding field of views of respectively 30\arcsec$\times$30\arcsec\ (blue channel) and 60\arcsec$\times$60\arcsec\ (red channel). Each pixel is a so-called "spaxel", which means that internally the light is dispersed spectrally (using a grating) over 16 pixels for each spaxel, providing an integral-field data cube for each observation, covering a total spectral bandwidth between 1500 and 3000\kms . The detectors consist of stressed, gallium-doped germanium photoconductors. 

For both \OIII\ lines and the \CII\ line the data was acquired in asymmetric chop mode with a fast mapping "AABAA" scheme \citep{Fischer2016}. The chop positions were determined based on the \her /PACS 160\mic\ continuum emission. A full chop throw of $\sim$ 10\arcmin\ was used to avoid chopping into emission. The map is a mosaic of 120 fields in a 30\arcsec\ raster. The total extent of the rectangular map is $\sim 5\arcmin \times 6\arcmin $ ($\sim$73\,pc $\times$ 87\,pc). For the \OIII\ line at 52\mic , {for which the point spread function (PSF) is the smallest}, the map was integrated twice with a (1.5, 0.5) pixel offset for better spatial sampling. The on-source integration times per point are summarised in Table~\ref{tab:obs}. Despite the fact that the \CII\ line was observed in parallel with \OIII\ 52\mic , the resulting on-source time is longer due to overlap of the larger field of view in the red channel. With the overheads for chopping and telescope moves, the whole map took about 150 min of wall-clock time at \OIII\ 52\mic\ and \CII\ and 75 min at \OIII\ 88\mic . For the weaker \OI\ line at 145\mic\ the symmetric chop mode was used with a $\sim 6\arcmin $ chop with multiple chop angles to avoid chopping into emission. Multiple fields were arranged around the existing data from PACS to extend the coverage to the required field of view ($\sim 5\arcmin \times 6\arcmin $). The wall-clock time for this map was about 11.5h.

The spectral resolution for the data and the full width half maximum (FWHM) of PSF for the data (taking into account the instrumental broadening) are summarised in Table~\ref{tab:obs}. The observations were obtained as part of SOFIA DDT project 75\_0016 and the open time proposal 05\_100.  The data has been taken on 8 SOFIA flights during 2 flight series between June 2016 and July 2017 (Cycles 4 and 5) during SOFIA Southern hemisphere deployments to Christchurch, New Zealand.

\begin{table}
    \centering
    \begin{tabular}{lccc}
    \hline
        Line & Integration  &  PSF  & Spectral \\
             & time per point         &  (FWHM) & resolution\\
             & (s)     & (arcsec) & (\kms )\\
    \hline
        \OIII\ 52\mic\ &  20   &  7.1   & 300 \\
        \OIII\ 88\mic\ &  10   &  10.0  & 490 \\
        \OI\ 145\mic\  &  1900 &  17.4  & 300 \\
        \CII\ 158\mic\ &  80   &  18.4  & 300 \\
    \hline
    \end{tabular}
    \caption{Characteristics of the FIFI-LS observations of 30Dor for the projects 75\_0016 and 05\_100.}
    \label{tab:obs}
\end{table}

\emph{Data reduction and flux calibration.}
Data reduction was carried out using the FIFI-LS data reduction pipeline \citep{Vacca2019}. The chop and nod cycles are first subtracted. Then, each spectrum is resampled to a regular wavelength grid with approximately eight pixels per FWHM of the instrument's spectral profile function at the observing wavelength. The spectra from the 25 spaxels are re-arranged back to their original 5 $\times$ 5 positions on the sky which differ slightly from the ideal 12\arcsec\ rectangular raster. 

A telluric correction is then applied and all data cubes are combined to get full maps with regular spatial sampling. At constant wavelength, spatial positions are interpolated locally with a polynomial surface fit to a rectilinear, regular spatial grid in right ascension and declination. The spatial sampling is 1\arcsec\ for the blue channel and 2\arcsec\ for the red channel. Flux calibration is established by observing calibration sources such as the planet Mars and measured spectra are compared to theoretical spectra, scaled to the distance between Mars and Earth. A full calibration across the wavelength range is done every flight series. The absolute flux calibration uncertainty is assumed to be 20\%, of which 10\% is relative uncertainty from FIFI-LS seen between flight series and different calibrators. The remaining uncertainties arise from atmospheric corrections and calibrator modelling uncertainties.

\emph{Atmospheric correction.}
Absorption of far-infrared light in the atmosphere due to water vapour or trace gases like ozone can have a significant impact on determining the true astronomical signal even at typical observing altitudes between 39,000 and 45,000 feet. Thus, telluric correction is critical to measure source flux. The main influencing factors for atmospheric transmission are flight altitude, source elevation and precipitable water vapour (PWV). The PWV is variable and could not be measured on board SOFIA at the time of observation. Predictions of the variability are available from weather forecasts but have not proven to be reliable.

The FIFI-LS pipeline uses a modelled transmission spectrum appropriate for the flight altitude and source elevation, with a standard value for water vapor overburden, as calculated with the ATRAN model \citep{Lord1992}. Based on a worst-case assumption of a PWV uncertainty of 50\%, the transmission uncertainty here is smaller than 5\% for all observed lines. After the spectral resampling of each nod subtracted data cube, the spectrum is divided by the convolved transmission spectrum for that cube (total integration time per cube is about 75s). This allows a correction for changing elevation and altitude on the time scale of minutes. This technique can be problematic near narrow atmospheric features since the convolution with the instrument's resolution significantly changes the shape of those features whereas the astronomical signal interacts with the unconvolved transmission profile of the atmosphere. However, there are no narrow lines at the specific observed line locations in 30Dor. 

\begin{figure*}
     \centering
      \includegraphics[trim=14cm 0mm 6mm 0mm, clip, width=8cm]{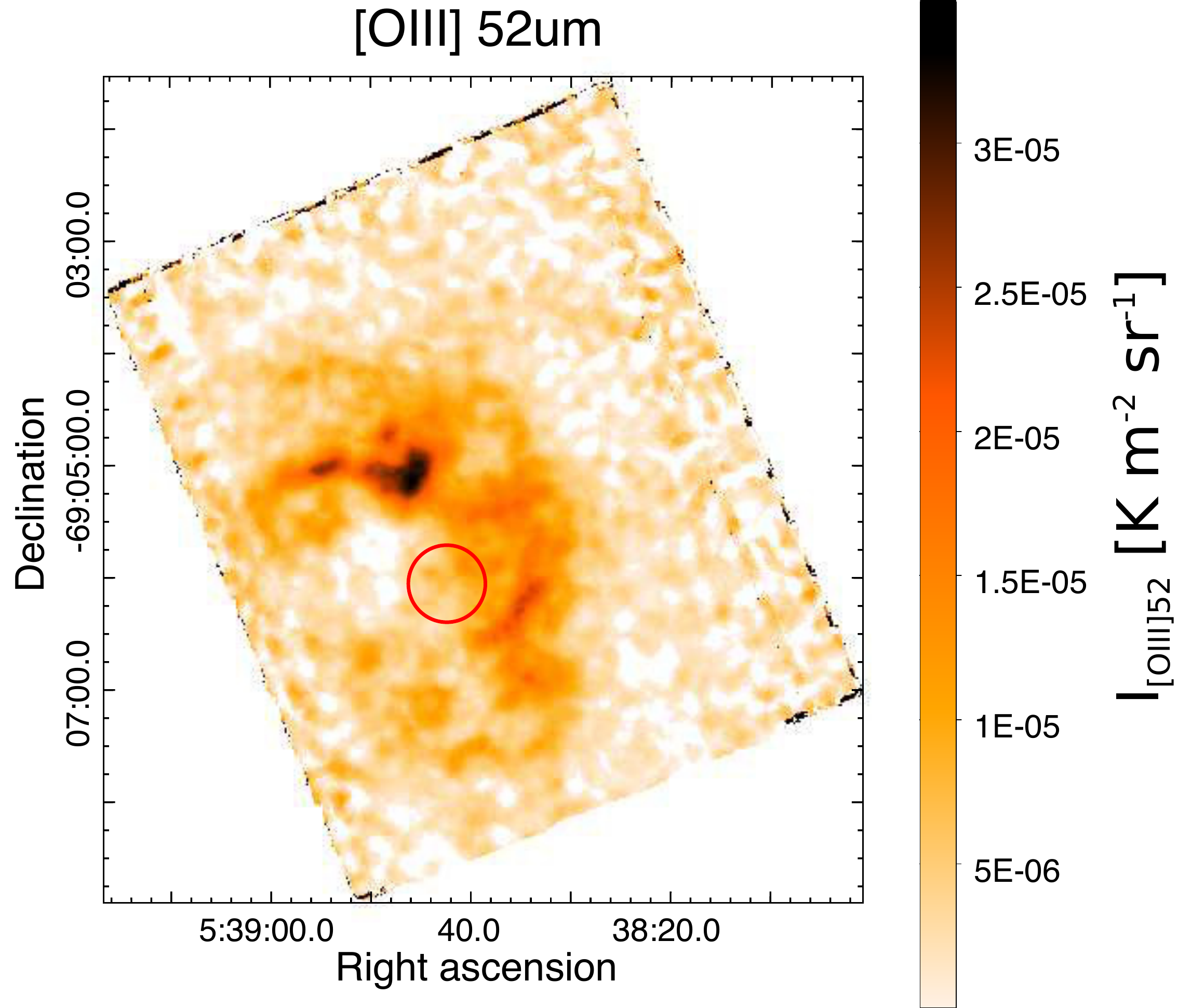}
      \includegraphics[trim=14cm 0mm 6mm 0mm, clip,width=8cm]{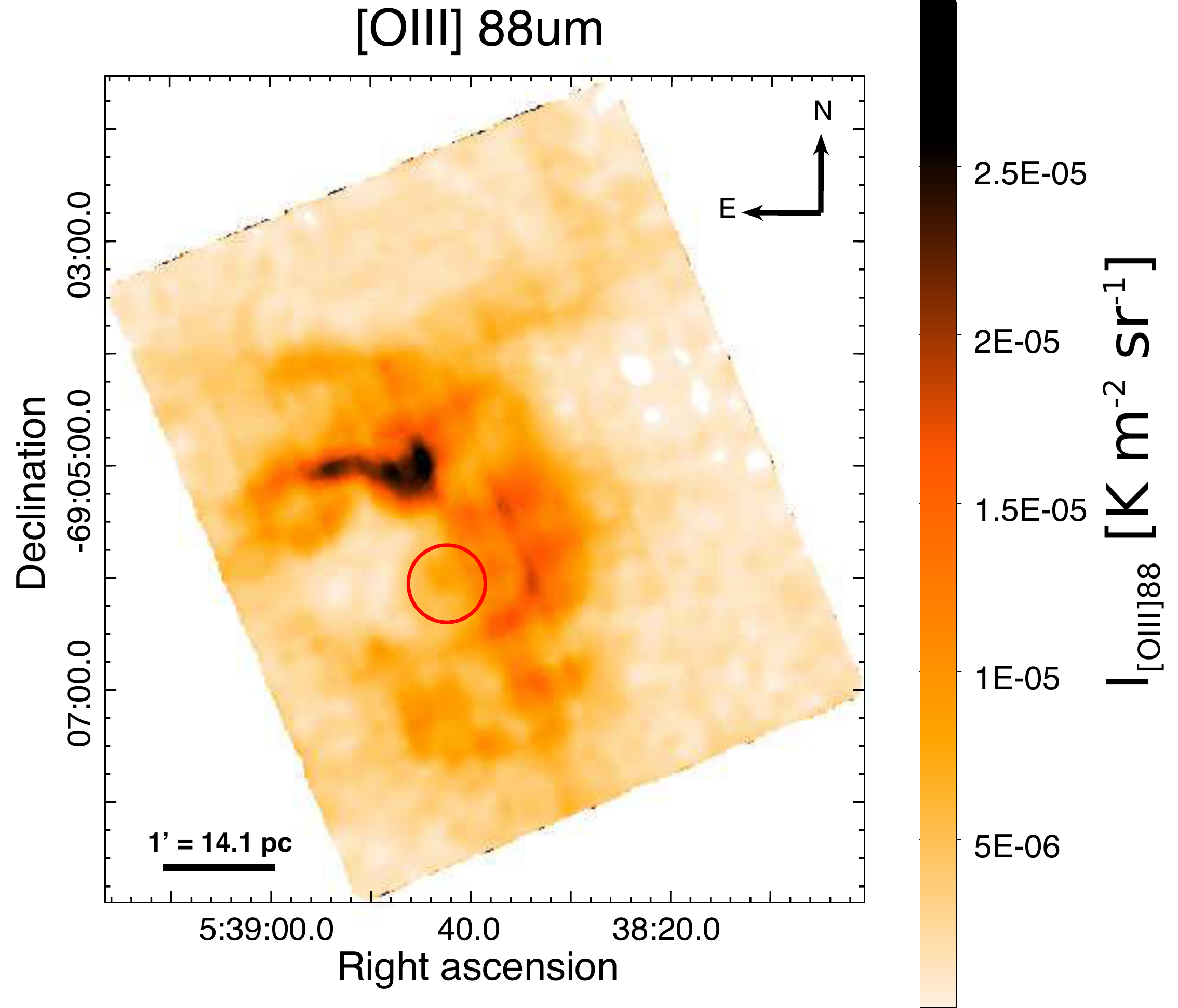}
      \includegraphics[trim=14cm 0mm 6mm 0mm, clip,width=8cm]{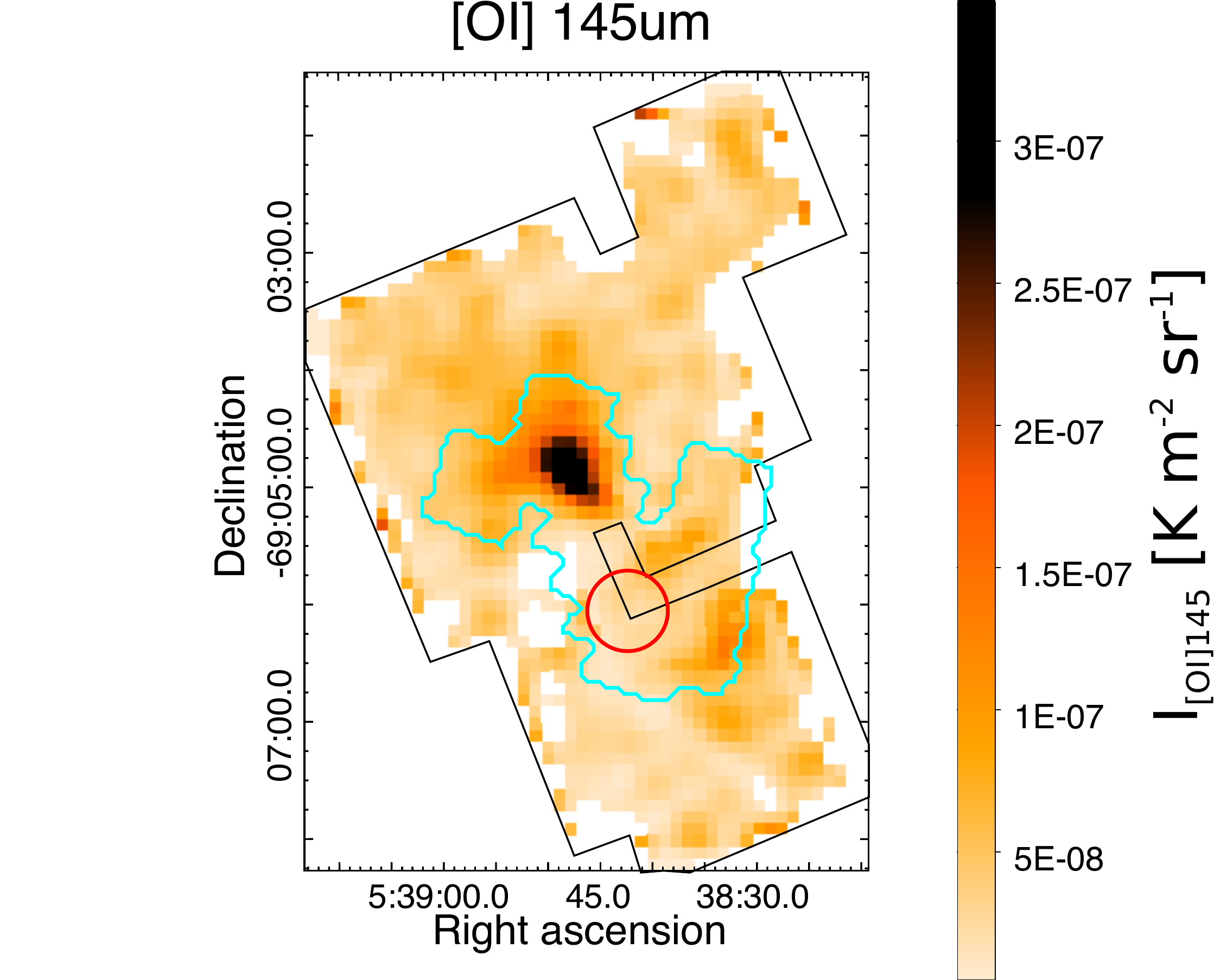}
      \includegraphics[trim=13.5cm 0mm 6mm 0mm, clip,width=8cm]{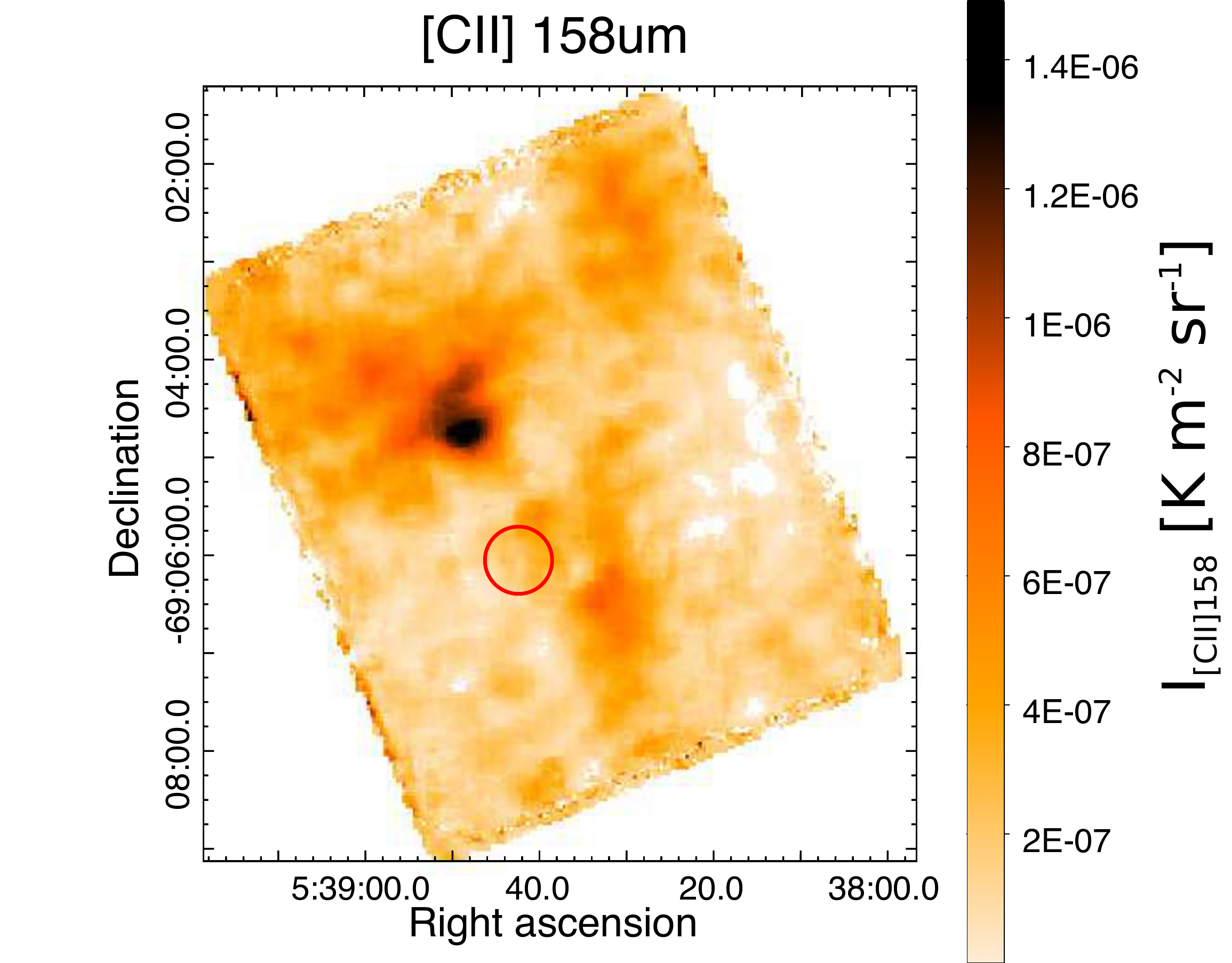}
            \caption{Intensity of the FIR fine structure lines \OIII\ at 52\mic , \OIII\ at 88\mic , \OI\ at 145\mic\ and \CII\ at 158\mic\ in \wmsqsr , observed by \sofia /FIFI-LS. The red circle has a 10\arcsec\ diameter and indicate the location of R136. For \OI\ 145\mic , the map is a combination of the FIFI-LS and PACS observations (see text). The outline of the PACS and FIFI-LS fields of view are indicated in cyan and black respectively. The white pixels in the maps are the ones for which the emission line fit has a a signal-to-noise ratio lower than 3. These are not taken into account in our analysis. We note the layered structure of the gas, with the peak of the ionised gas (\OIII ) being closer to R136 than the peak of the neutral gas (\CII\ and \OI ). The latter also corresponds to the peak of the molecular gas mass surface density as determined in Section~\ref{sec:pdr_H2}.}
       \label{fig:SOFIA}
    \end{figure*}

\emph{Flux measurement.}
We use \textsc{Fluxer}\footnote{\url{http://www.ciserlohe.de/fluxer/fluxer.html}} to measure the flux from the combined and calibrated cubes.
For each pixel, the continuum is modeled with a zeroth order polynomial and the emission line is fitted using a gaussian function. The resulting emission maps for \OIII\ at 52 and 88\mic , \CII\ at 158\mic\ and \OI\ at 145\mic\ are presented in Figure~\ref{fig:SOFIA}.

\emph{Combination of PACS and FIFI-LS observations for \OI\ 145\mic .}
As the \OI\ 145\mic\ fine structure line requires the longest integration time with FIFI-LS, we have maximised the total coverage by not duplicating the PACS observations. We have ensured partial overlap between the maps for cross-calibration purposes. As a result, a small region in the central part of 30Dor was not observed by FIFI-LS in \OI\ 145\mic , as visible in Figure~\ref{fig:SOFIA}. To recover the full map, we combine the FIFI-LS and PACS maps as follows. We first check the agreement between the relative flux calibrations of the FIFI-LS and PACS observations for the areas that were observed by both instruments (equivalent to approximately 20 FIFI-LS beams). They agree at the $\sim$ 10\% level on 12\arcsec\ beam size once both maps are convolved to the same spatial resolution. In view of the $\sim$20\% absolute calibration of PACS and FIFI-LS we can then establish that the relative flux calibrations between both sets of observations are in agreement and we therefore simply match the FIFI-LS and PACS maps on the same 2\arcsec $\times$ 2\arcsec\ pixel grid. We assign the observed FIFI-LS (respectively PACS) flux to pixels that only have flux observation above a signal-to-noise ratio (SNR) of 3 with FIFI-LS (respectively PACS), and the SNR-weighted mean flux of the FIFI-LS and PACS fluxes for pixels that are covered by both instruments.

\subsection{Archival data}
The \her /PACS \citep{Poglitsch2010} spectroscopic data of 30Dor have been presented in \citet{Chevance2016}, and we refer to \cite{Cormier2015} for the full description of the PACS observations and data reduction. We use here only the maps of \OI\ at 145\mic\ to complement the field of view covered by FIFI-LS, as described in Section~\ref{sec:obs_FIFI}.

One of the parameter of the PDR model is the total extinction in the $V$ band, \Avmax , which scales with the depth of the PDR layer. Ratios of lines originating from the same depth in a PDR cloud are not sensitive to the size of the cloud (e.g. \CII\ and \OI ; see also Section~\ref{sec:pdr_results}). In addition to the PDR tracers observed by FIFI-LS, tracers originating from different depths into the cloud are therefore necessary to constrain the total depth of the cloud \Avmax . We use observations of the CO(1-0) from Mopra \citep{Wong2011} at a spatial resolution of 42\arcsec , which we also use to estimate the mass surface density of the CO-bright molecular gas (Section~\ref{sec:pdr_H2}). For the atomic gas, we use \HI\ observations of the LMC from \cite{Kim2003} at a spatial resolution of 1\arcmin .

In order to calculate the FIR luminosity, we use the \her /PACS and SPIRE maps of the Large Magellanic Cloud at 100, 160 and 250\mic , first published in \cite{Meixner2013} as part of the HERITAGE project. We also make use of the observations of 30Dor obtained as part of the \spit\ \citep{Werner2004} Legacy program "Surveying the Agents of a Galaxy's Evolution" (SAGE; \citealt{Meixner2006}). We used the four channels of IRAC \citep{Fazio2004} at 3.6, 4.5, 5.8 and 8.0\mic\ and MIPS \citep{Rieke2004} observations at 24 and 70\mic . As the MIPS 24\mic\ map is saturated in several pixels, we use the IRS spectra (\citealt{Indebetouw2009}; see also Sect.~\ref{sec:spitzer}) to calculate the 24\mic\ synthetic photometry in the MIPS 24 bandpass and compare to the original map. These are in excellent agreement in parts where the \spit /MIPS map is not saturated. We apply the dust SED model of \citet[AC composition]{Galliano2011}  to these data, as described in \citet{Chevance2016}. Since our analysis here is limited by the spatial resolution of the CO(1-0) observations (i.e. 42\arcsec ), we use the photometric bands of SPIRE up to 250\mic\ to constrain the SED model in order to maximise the accuracy on the dust mass (this contrasts with \citet{Chevance2016}, where only the FIR luminosity was used, and obtained at a good accuracy with the photometric bands up to 160\mic , which lead to a resolution of 12\arcsec ).

\section{Empirical diagnostics}
\label{sec:lines}

\subsection{\CII /\OIII\ ratio}

In metal-rich environments, \CII 158\mic\ and \OI 63\mic\ lines are the dominant FIR emission lines. However, this is often not the case in dwarf galaxies, at lower metallicity (e.g. \citealt{Hunter2001, Cormier2015}), where \OIII\ lines become more prominent.

\begin{figure}
     \centering
      \includegraphics[trim=8cm 0mm 0mm 0mm, clip, width=8cm]{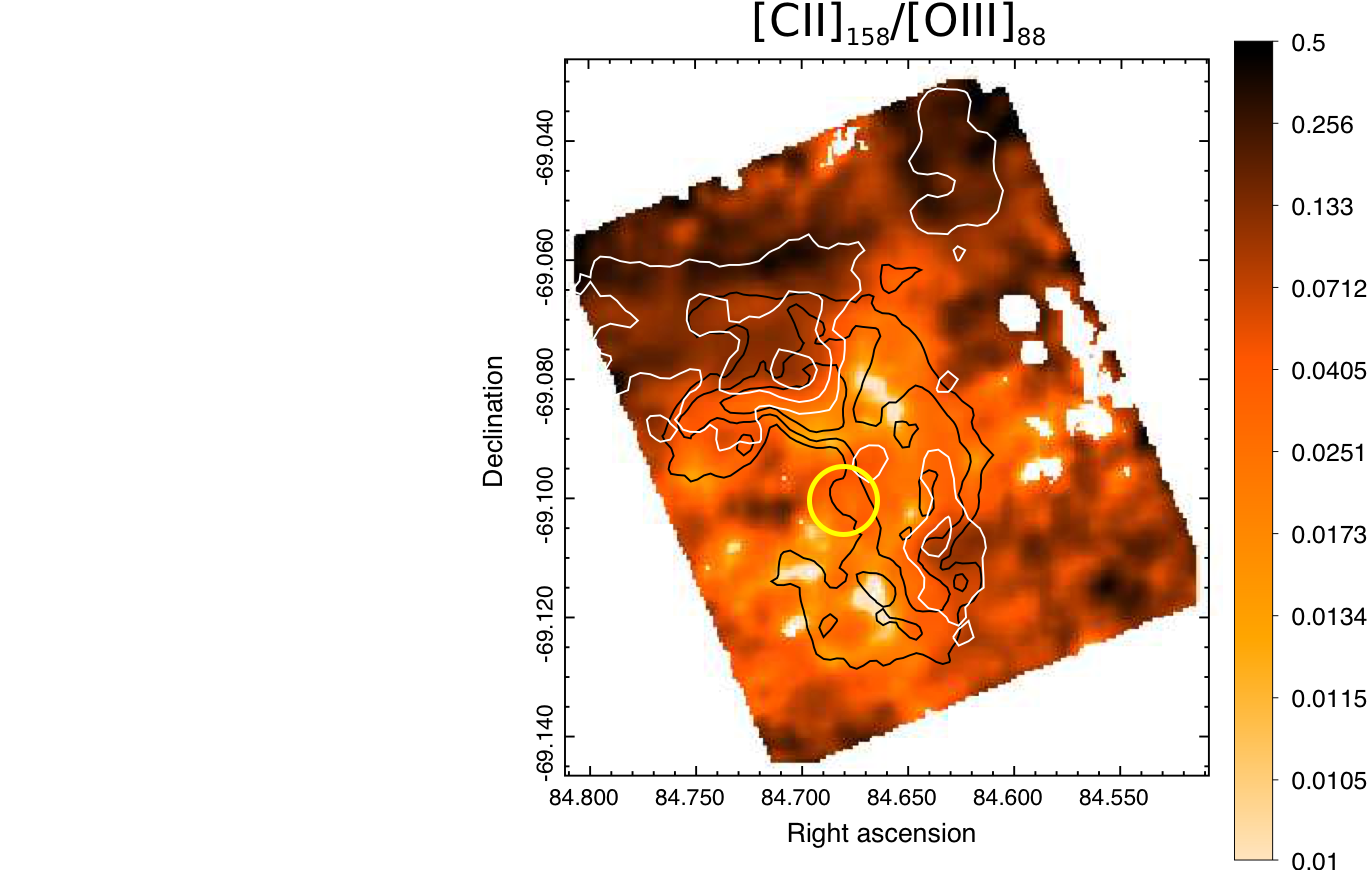}
       \caption{Ratio of the FIR fine structure lines \CII\ 158\mic\ / \OIII\ 88\mic\ at 18\arcsec\ ($\sim$4\,pc) resolution. We note that the color bar is in logarithmic scale. The black contours represent the \CII\ line intensity and the white contours represent the \OIII\ 88\mic\ line intensity. The yellow circle indicates the location of R136. The ratio \CII /\OIII\ 88\mic\ is below 1 in the entire area covered by FIFI-LS, revealing a porous medium bathed in strongly ionising radiation field tens of parsecs away from the excitation source R136.}
       \label{fig:CIIOIII}
    \end{figure}

In 30Dor, we now extend the analysis of \citet{Chevance2016}, covering projected distances from the centre of the excitation source R136 only up to $\sim$40\,pc, to the entire FIFI-LS map. We show that the emission from the \OIII\ lines (both at 52\mic\ and 88\mic ) is relatively extended, such that the \OIII 88\mic\ line is brighter than \CII\ in the entire covered area, up to a projected distance of $\gtrsim$ 60\,pc from the centre of R136 (see Figure~\ref{fig:CIIOIII}). The ratio \CII /\OIII 88\mic\ shows low values (between $\lesssim 10^{-2}$ and $\sim 0.5$ at the resolution of $\sim$\,4\,pc achieved by the FIFI-LS observations), with the lowest values reached around the peaks of the ionised gas probed by the \OIII\ lines. The ratio \CII /\OIII 52\mic\ ranges between $\lesssim 10^{-2}$ and $\sim 0.3$.

These low values reached by the ratio \CII /\OIII\ (even at large distances from the centre of R136) highlight the fact that high energy photons (the ionisation potential of O$^+$ is 35.1 eV) can travel over long distances before interacting with the ISM, revealing the high level of porosity of low-metallicity environments. As a result of this porosity of the ISM, R136 is indeed expected to be the dominant source of ionisation in 30Dor \citep{Kawada2011, Lopez2011}.

\subsection{Density of the ionised gas}

The ratio of the \OIII\ fine-structure cooling lines is a good probe of the electron density in the highly ionised gas. The \OIII\ 52\mic\ line was unobservable with \her\ in the LMC so this diagnostic could not easily be evaluated for star forming regions of the LMC before \sofia . The ratio of the \SIII\ lines at 18\mic\ and 33\mic , commonly observed by \spit , is sensitive to high gas densities (the critical densities of these lines are respectively $1.5\times10^4$ \cm\ and $4.1\times10^3$ \cm ), which provides only limited constraints in the (low) electron density regime of 30Dor ({\citealt{Indebetouw2009} estimate an electron density of 300-600\,\cm\ using this ratio}; see also \citealt{Chevance2016}). The critical density of the \OIII-emitting material ($n_{\rm crit,\rm{OIII88}} = 5.1 \times 10^2$\,cm$^{-3}$; $n_{\rm crit,\rm{OIII52}} =3.6\times10^3$\,cm$^{-3}$) is lower than those of classical dense \HII\ regions, allowing us to probe the low-density ionised gas.

\begin{figure*}
     \centering
      \includegraphics[trim= 65mm 0mm 15mm 0mm, clip, width=7.0cm]{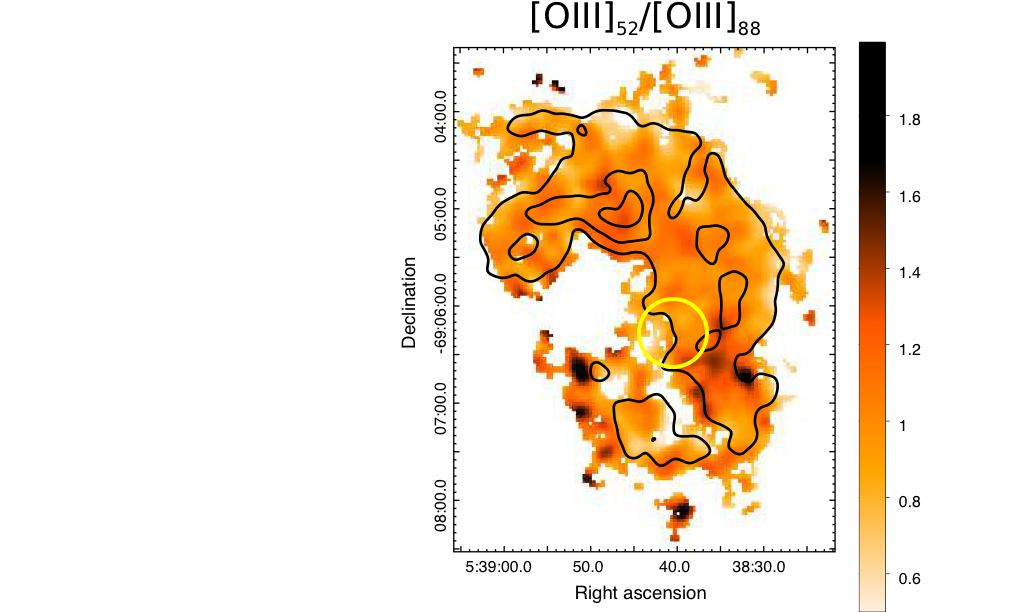}
      \includegraphics[trim= 0mm 0mm 0mm 0mm, clip, width=9.5cm]{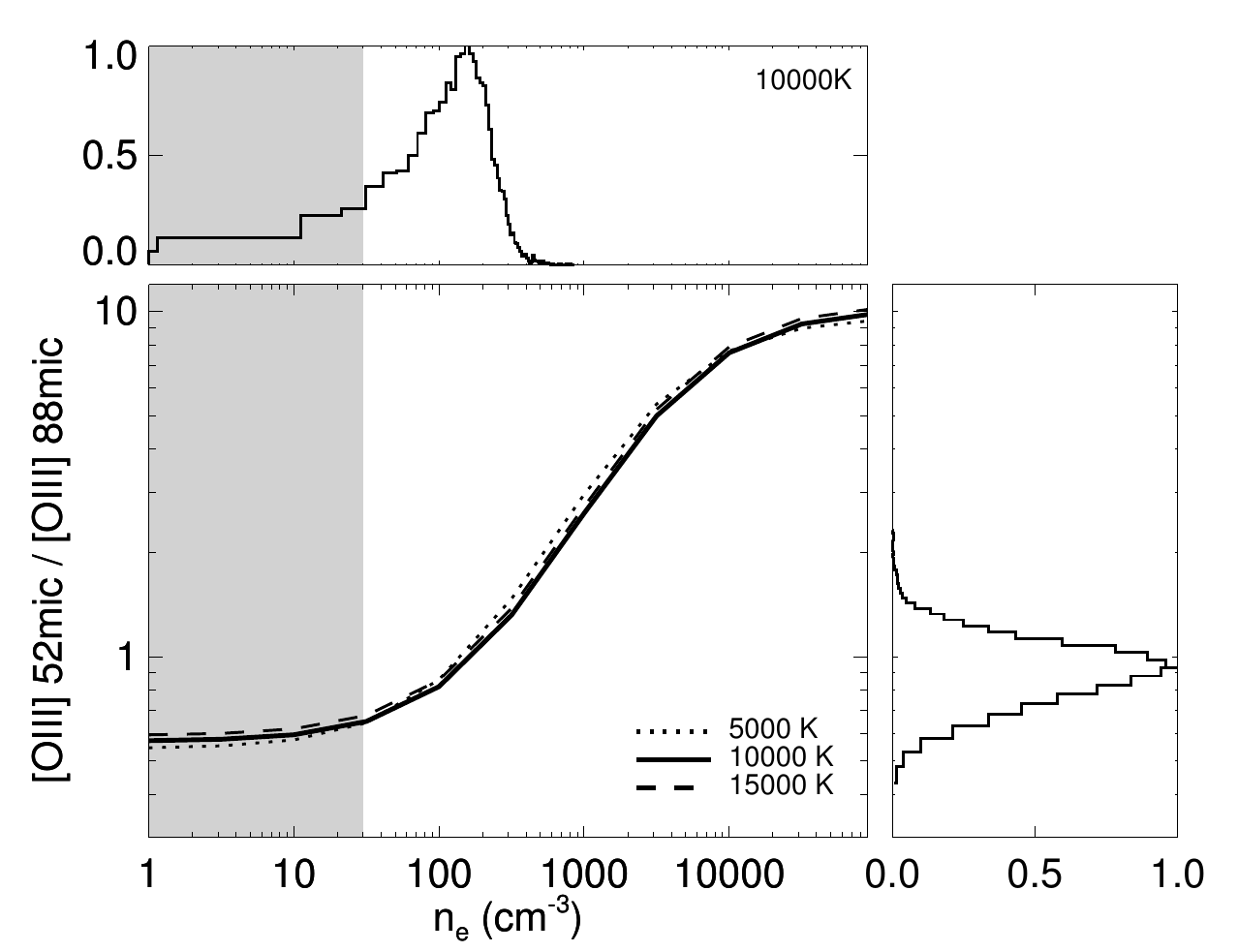}
       \caption{\textit{Left:} Map of the observed ratio \OIII\ 52\mic / \OIII\ 88\mic\ (restricted to the pixels with SNR > 5). The black contours represents the \OIII\ 88\mic\ intensity. The yellow circle indicates the location of R136. 
       \textit{Right:} Theoretical ratio \OIII\ 52\mic / \OIII\ 88\mic\ at the temperatures of 5\,000 (dotted line), 10\,000 (solid line) and 15\,000\,K (dashed line) as a function of the electron density $n_e$. The normalised histogram of the observed values of the ratio \OIII\ 52\mic / \OIII\ 88\mic\ is shown on the right. The normalised histogram of the inferred densities (using the theoretical prediction at 10\,000\,K) is shown on the top. For the lowest value of the observed ratio ($\lesssim$ 0.65), only an upper limit can be determined for the density. The values in the grey area should therefore be considered as upper limits at $\sim 30$\cm .
       }
       \label{fig:OIII}
    \end{figure*}

The ratio map of \OIII 52\mic /\OIII 88\mic , along with the theoretical variations of this ratio as a function of the electron density and temperature are shown in Figure~\ref{fig:OIII}. The theoretical ratio depends mostly on the density and only slightly on the temperature. We choose a typical temperature of 10\,000\,K for the ionised gas around a star-forming region to determine the density constrained by our observations. In our FIFI-LS observations of 30Dor, the ratio \OIII 52\mic /\OIII 88\mic\ ranges between 0.48 and 1.84 at a resolution of $\sim 10\arcsec $, leading to a maximum electron density of a few $10^2$ cm$^{-3}$ (up to $\sim 500$ \cm ). Where the ratio is lower than $\sim$ 0.65 (typically at large distances from the centre of R136), we can only infer an upper limit of $\sim$ 30 cm$^{-3}$ for the electron density.

These values are on average larger than the density range of 10-100 \cm\ probed in the vicinity of the peak of the \OIII\ lines in \cite{Chevance2016}, using the ratio \NII\ 122\mic\ / \NII\ 205\mic\ at a lower spatial resolution of $\sim$ 20\,pc. This is not unexpected since the critical densities of the \NII\ lines (48\,\cm\ and 310\,\cm ) are lower than that of the \OIII\ lines, and lower densities are measured on average at lower resolution (with a maximum of $\sim 350$ \cm\ measured from the \OIII\ line ratio at 20\,pc resolution). The densities measured using the \OIII\ line ratio is however broadly consistent with the values measured by \cite{Pellegrini2011} from \SII\ observations, although we only probe the centre of the region. The \OIII , \SIII , \SII\ and \NII\ lines therefore seem to be tracing different phases of the ISM, as expected from their different excitation potentials (35.1\,eV, 23.3\,eV, 10.4\,eV and 14.5\,eV for O$^{++}$, S$^{++}$, S$^{+}$ and N$^{+}$, respectively) and critical densities. The range of physical conditions probed by these different tracers is investigated in Polles et al. (in prep.).

\section{PDR modeling}
\label{sec:PDR}

In the proximity of a star forming region in low-metallicity environments, the strong FUV radiation from the nearby massive stars combined with the reduced metal and dust abundances results in a shift of the \Cp /\Co /CO transition deeper into the gas clouds. This effect induces a smaller CO core, surrounded by a larger shell of \Co\ and \Cp\ which can potentially encompass an important reservoir of \HH\ \citep[e.g.][]{Wolfire2010, Bolatto2013}. The \HH\ molecule is self-shielded at a smaller column density relative to that for the CO molecule, which makes the photodissociation rate of CO larger than that of H$_2$ at the cloud surface. 

In this section, we show how detailed PDR modeling can be used to determine the physical conditions of the neutral gas and we compute the depths at which \HH\ and CO form. This model, constrained by observations, allows us to calculate the total mass of molecular gas.

\subsection{Physical conditions of the PDR}
\label{sec:pdr_results}
We use the Meudon PDR model \citep{LePetit2006, LeBourlot2012, Bron2014} to calculate the mass surface density of molecular gas in 30Dor. This model computes the atomic and molecular structure of interstellar clouds in a 1D stationary plane parallel slab of gas and dust illuminated by a radiation field (from UV to radio) arising from one or both sides. We use a constant pressure equation of state throughout the cloud. 

\begin{figure*}
     \centering
      \includegraphics[trim= 16cm 0mm 2mm 0mm, clip, height=5.8cm]{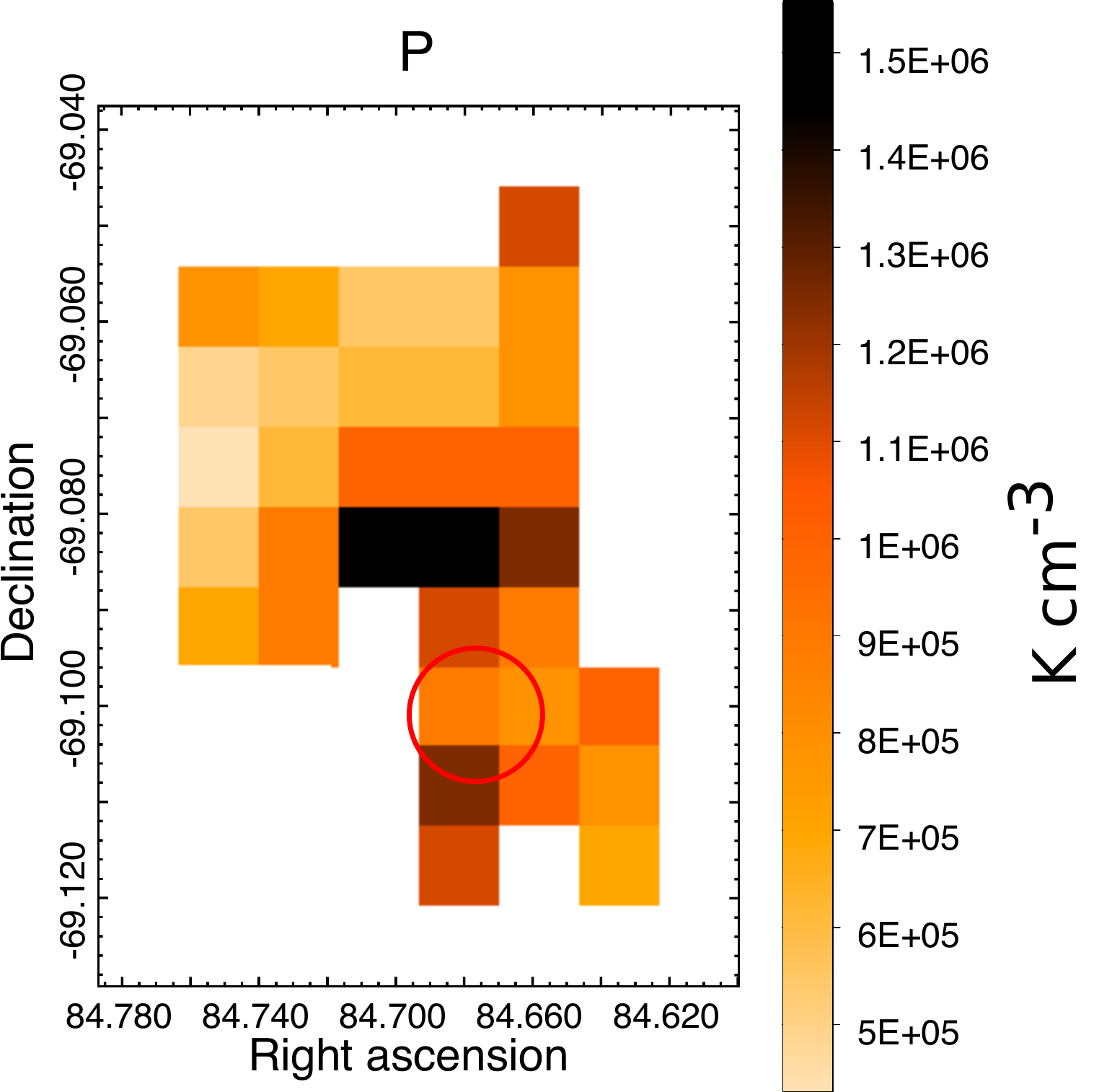}
      \includegraphics[trim= 16cm 0mm 10mm 0mm, clip, height=5.8cm]{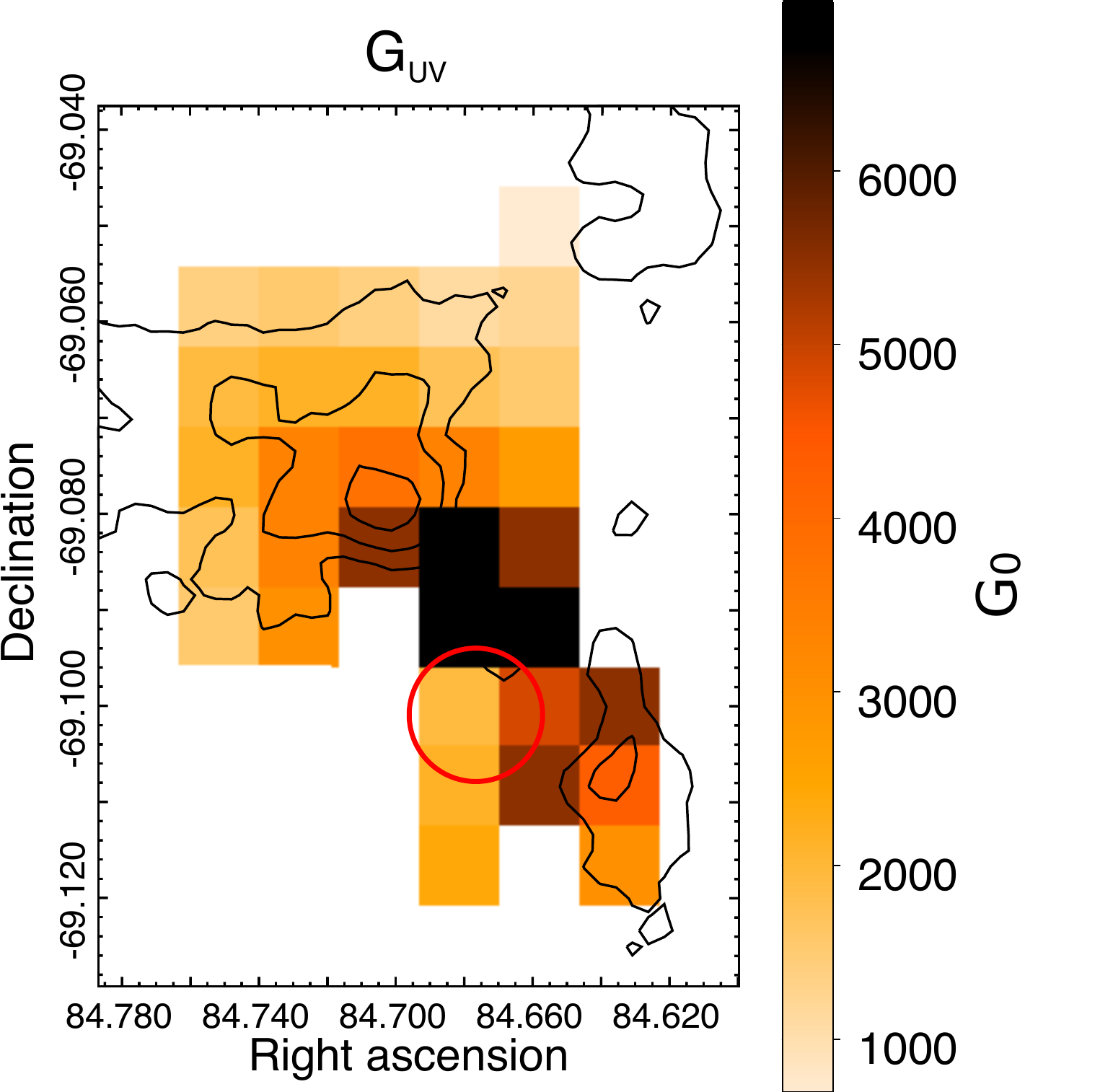}
      \includegraphics[trim= 16cm 0mm 15mm 0mm, clip, height=5.8cm]{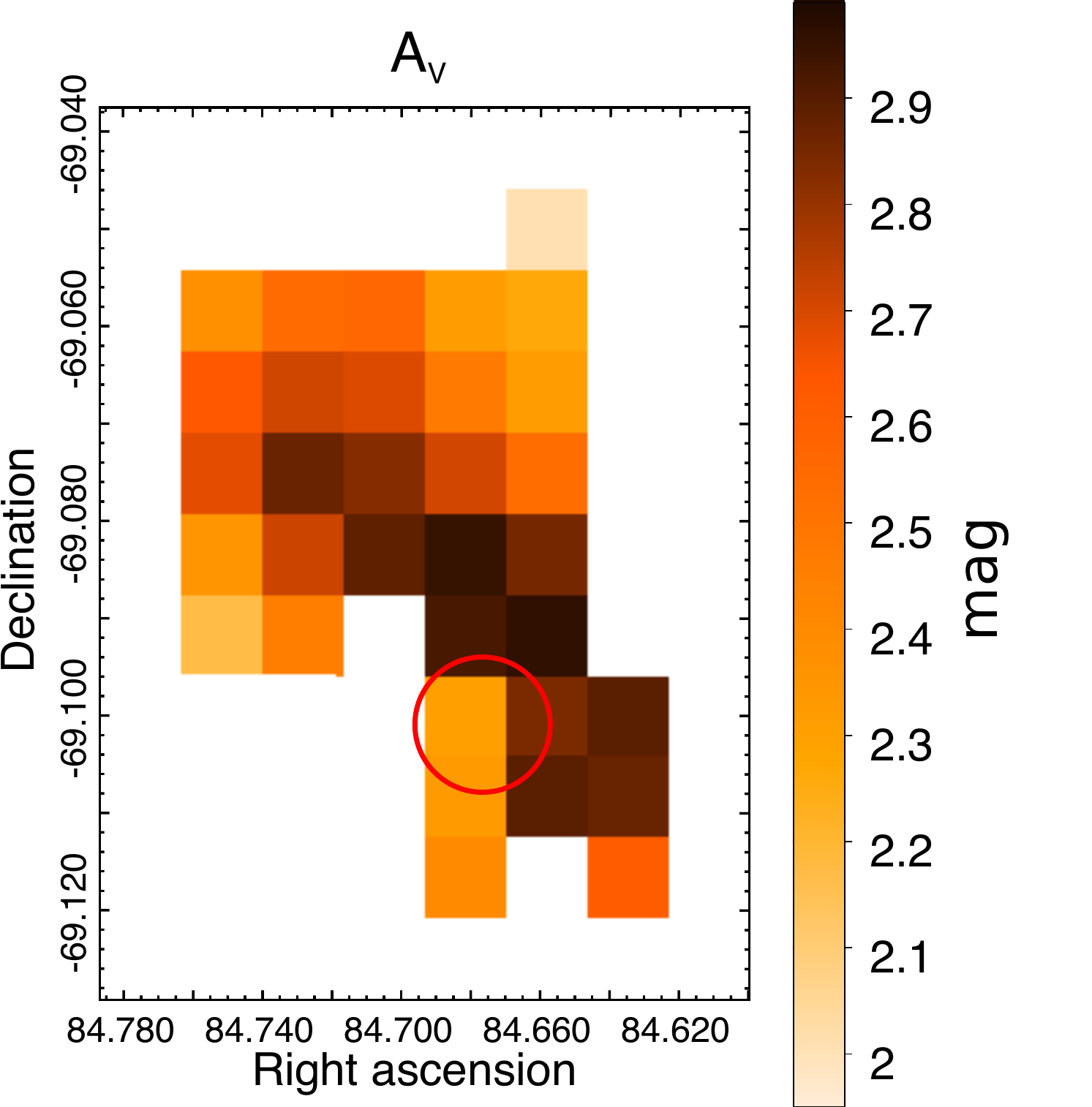}
       \caption{Constrained parameter of the Meudon PDR model, using the observed ratios $(\text{\OI}_{145} + \text{\CII}_{158})/{\fir}$, ${\text{\OI}_{145}}/{\text{\CII}_{158}}$ and ${\text{\CII}_{158}}/{\text{CO(1-0)}}$, at a resolution of 43\arcsec . The spatial distributions of the pressure $P$ (\textit{left}; in K \cm ), the incident radiation field (\textit{centre}; in units of the Mathis field) and the total visual extinction \Avmax\ (\textit{right}; in magnitudes) are presented, along with the location of the SSC R136 (red circle). We overplot the \CII\ intensity as black contours in the central panel to guide the eye.}
       \label{fig:PDRmodel}
    \end{figure*}

We use the method described in \citet{Chevance2016} to find the PDR model parameters that provide the best fit to the observed line ratios $(\text{\OI}_{145} + \text{\CII}_{158})/{\fir}$, $\text{\OI}_{145}/{\text{\CII}_{158}}$ and ${\text{\CII}_{158}}/{\text{CO(1-0)}}$, at a spatial resolution of 43\arcsec\ (determined by the CO(1-0) observations). The parameters we fit for are the incident radiation field on one side of the cloud, \go , the pressure in the gas, $P$, and the total visual extinction of the model, \Avmax . We fit for these three parameters within each 30\arcsec\ pixel. The resulting spatial distributions of these three parameters are presented in Figure~\ref{fig:PDRmodel}. The incident radiation field on the PDR, \go , ranges between $7\times10^2$ and $7\times10^3$ (in units of the standard radiation field observed in the solar neighborhood, where the integrated energy density between 911.8 \AA\ to 2400 \AA\ is G$_0 = 6.8 \times 10^{-14}$ erg \cm ; \citealt{Mathis1983}). We note that we observe a minimum of \go\ in the direction of R136. This indicates that the surface of the PDR is likely to be far away from the cluster in that region, and therefore the received flux by the PDR is low (see Section~\ref{sec:variation}). The measured pressure, $P$, ranges between $4.3\times10^5$ and $1.6\times10^6$ K \cm . \Avmax\ ranges between 2 and 3 mag. By comparing the predictions of the constrained model for each pixel to the \CII\ observations, we measure the area filling factor, \ff , similarly to \citet{Chevance2016}. We find \ff $\sim 1-6$, which means that typically several clouds are present along the line of sight.

Given the lower resolution at which this analysis is performed (set by the CO data), the results for \go , $P$, and \ff\ measured here are consistent with the results of \citet{Chevance2016}. Our measurements of \go\ and $P$ are also in close agreement with the values constrained at a similar spatial resolution by \cite{Lee2019}, using the Meudon PDR model, and using the \CI\ line at 370\mic\ instead of the CO(1-0) line. If we exclude the constraint provided by the ratio $\frac{\text{\CII}_{158}}{\text{CO(1-0)}}$, it is possible to constrain the PDR parameters \go\ and $P$ at a higher resolution of $\sim$ 18\arcsec , limited by the FIFI-LS \CII\ observations, instead of 43\arcsec\ as presented here. In this case, the values and spatial distributions of \go\ and $P$ using the FIFI-LS observations are in close agreement with the results of \citet{Chevance2016} using the PACS observations, but this does not allow us to constrain \Avmax , necessary for the determination of the total molecular gas mass. The reason is that \CII\ and CO come from different depths into the PDR cloud, which makes the ratio \CII /CO sensitive to \Avmax , contrary to ratios involving only \CII , \OI , and \Lfir\ (originating from similar parts of the PDR cloud). We note that \Avmax\ constrained here using the ratio \CII /CO(1-0) is slightly larger than \Avmax\ constrained by \citet{Lee2019} ($\sim$ 1.5-2\,mag), relying mostly on the ratio \CII / \CI . This is not surprising given the expected structure of PDR regions, where CO lines are emitted deeper into the clouds than \CI\ lines, and as already shown in fig.~17 of \citet{Chevance2016}. The average \Avmax\ for an ensemble of clouds is therefore likely to be smaller for \CI\ emitting clouds than for CO emitting clouds. We note however that this small uncertainty on \Avmax\ only slightly affects the predicted \HH\ mass and the results presented in Section~\ref{sec:pdr_H2}, since the transition between H and \HH\ occurs closer to the surface of the cloud (\Av\ $\lesssim 1$).

Recent observations of several regions of the LMC, including 30Dor, have also been made in \CII\ with the German REceiver for Astronomy at Terahertz frequencies (GREAT) and in \OI\ at 63\mic\ and 145\mic\ with upGREAT \citep{Risacher2016}. These observations are presented by \cite{Okada2019}, who combined them with the CO transitions from $J$ = 2--1 to $J$ = 6--5, \CI\ $^3 \rm P _1 - ^3 \rm P _0$, and \CI\ $^3 \rm P _2 - ^3 \rm P _1$, to constrain the parameters of the KOSMA-$\tau$ PDR model \citep{Storzer1996, Rollig2006, Rollig2013}, at 30\arcsec\ resolution. We note that they use an isochoric PDR model (with a constant gas density through a given PDR cloud), while we use an isobaric model (with a constant pressure through a given PDR cloud). We show in \citet{Chevance2016} that the choice of an isochoric versus isobaric PDR model has little effect on the determination of \go\ (at the $\sim$ 20\% level). However, the CO lines are better reproduced when using an isobaric model, which is what is driving our choice for an isobaric model here. \cite{Okada2019} find \go\ between $\sim 10^2$ and $10^{5.5}$ and densities $n$ between $\sim 10^{3.5}$ and $10^7$ \cm\ throughout 30Dor, with similar distributions as in Figure~\ref{fig:PDRmodel}. The values of $n$ \citep{Okada2019} and $P$ (this study) are not straightforward to compare, but we note that the values of \go\ derived with the KOSMA-$\tau$ PDR model reach much higher values than that constrained with the Meudon PDR model, which seem inconsistent with the emitted stellar radiation derived in \citet{Chevance2016}. These could result from the degeneracy between a high $n$-high \go\ and a low $n$-low \go\ solution as noted by \cite{Okada2019}.

\subsection{Total \HH\ reservoir}
\label{sec:pdr_H2}

\begin{figure*}
     \centering
     \hspace{-0.3cm}
      \includegraphics[trim= 16cm 0mm 15mm 0mm, clip, height=5.8cm]{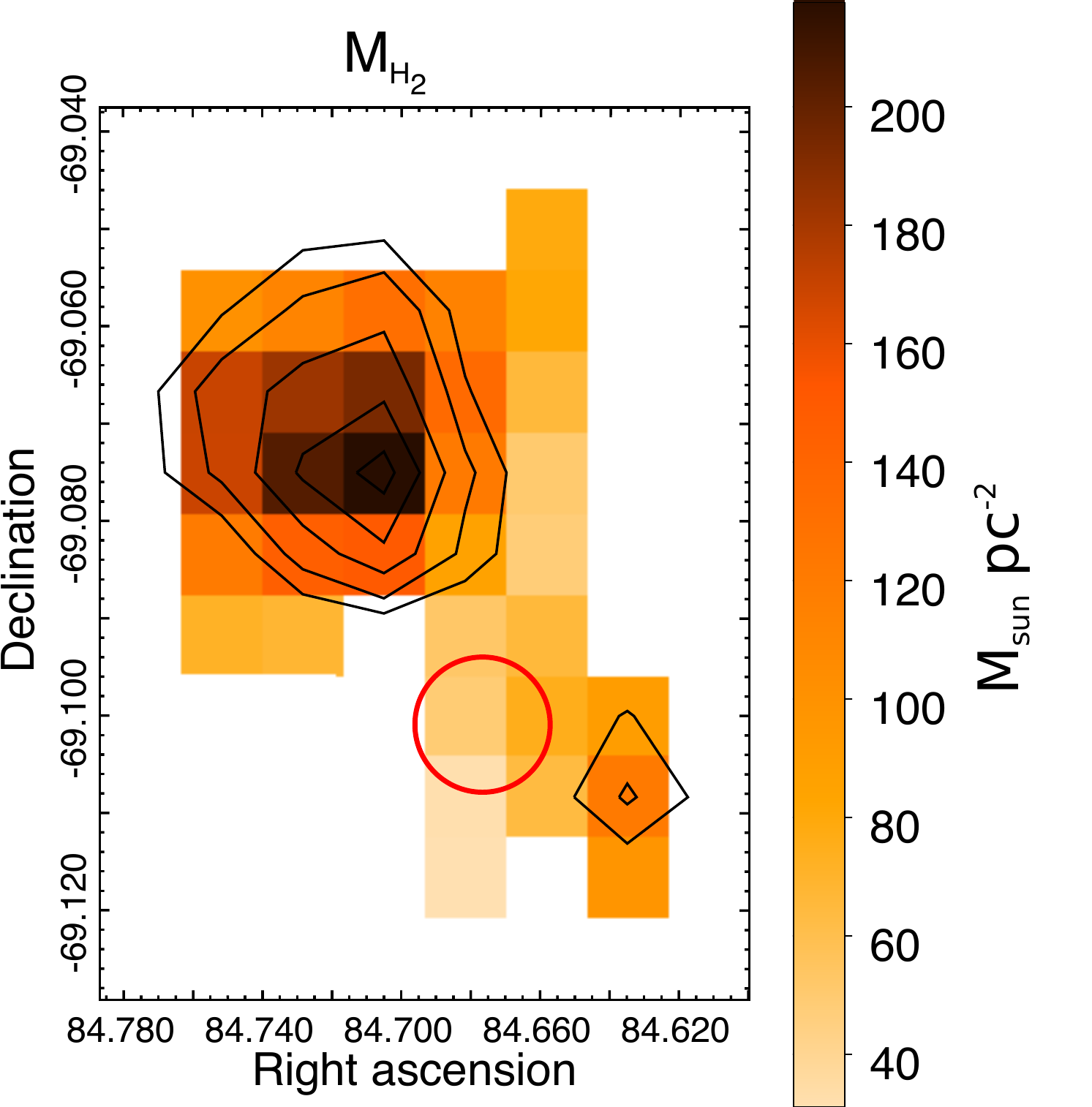}
      \includegraphics[trim= 6mm 3mm 50mm 40mm, clip, height=5.8cm]{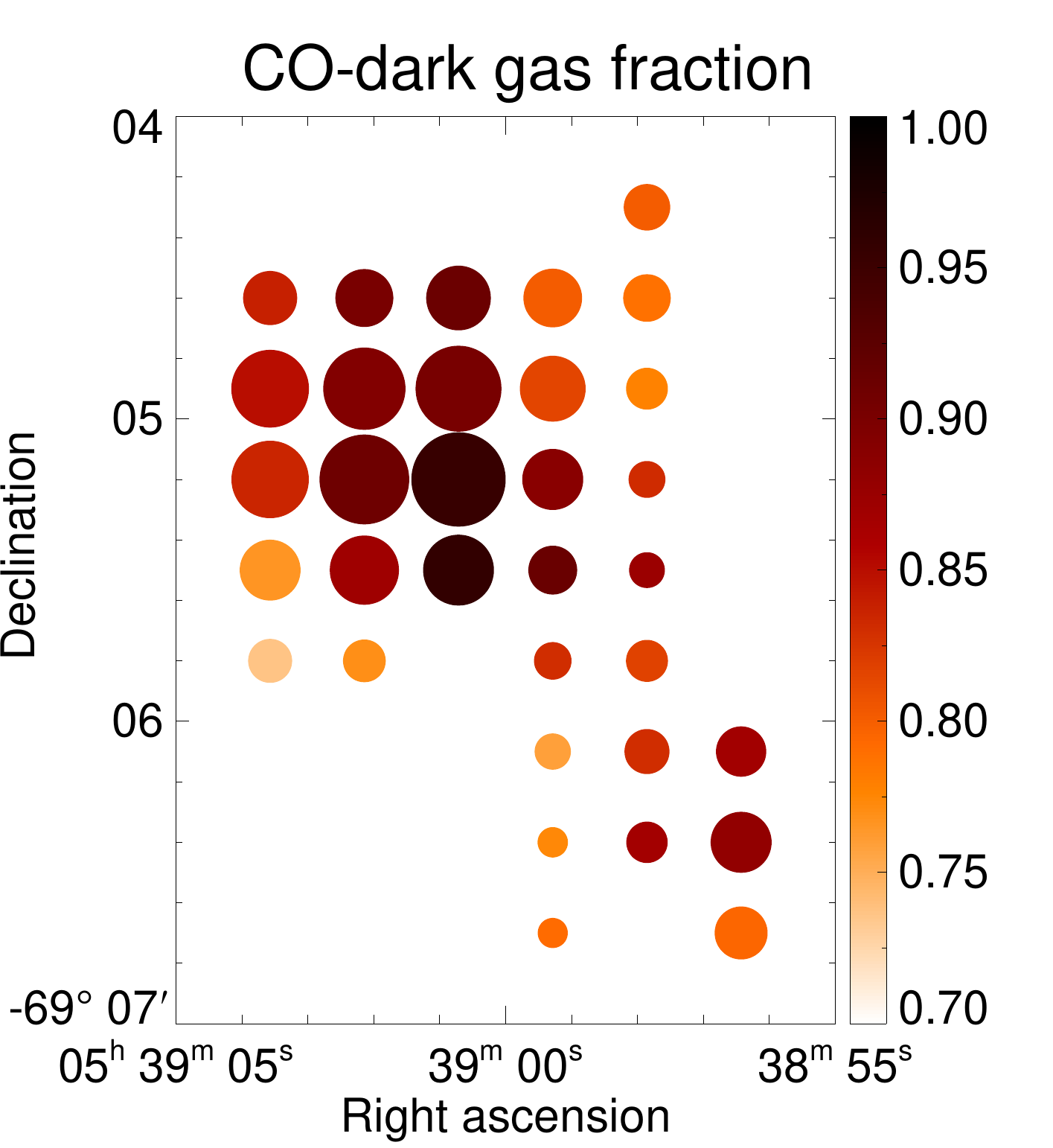}
      \includegraphics[trim= 16cm 0mm 25mm 0mm, clip, height=5.8cm]{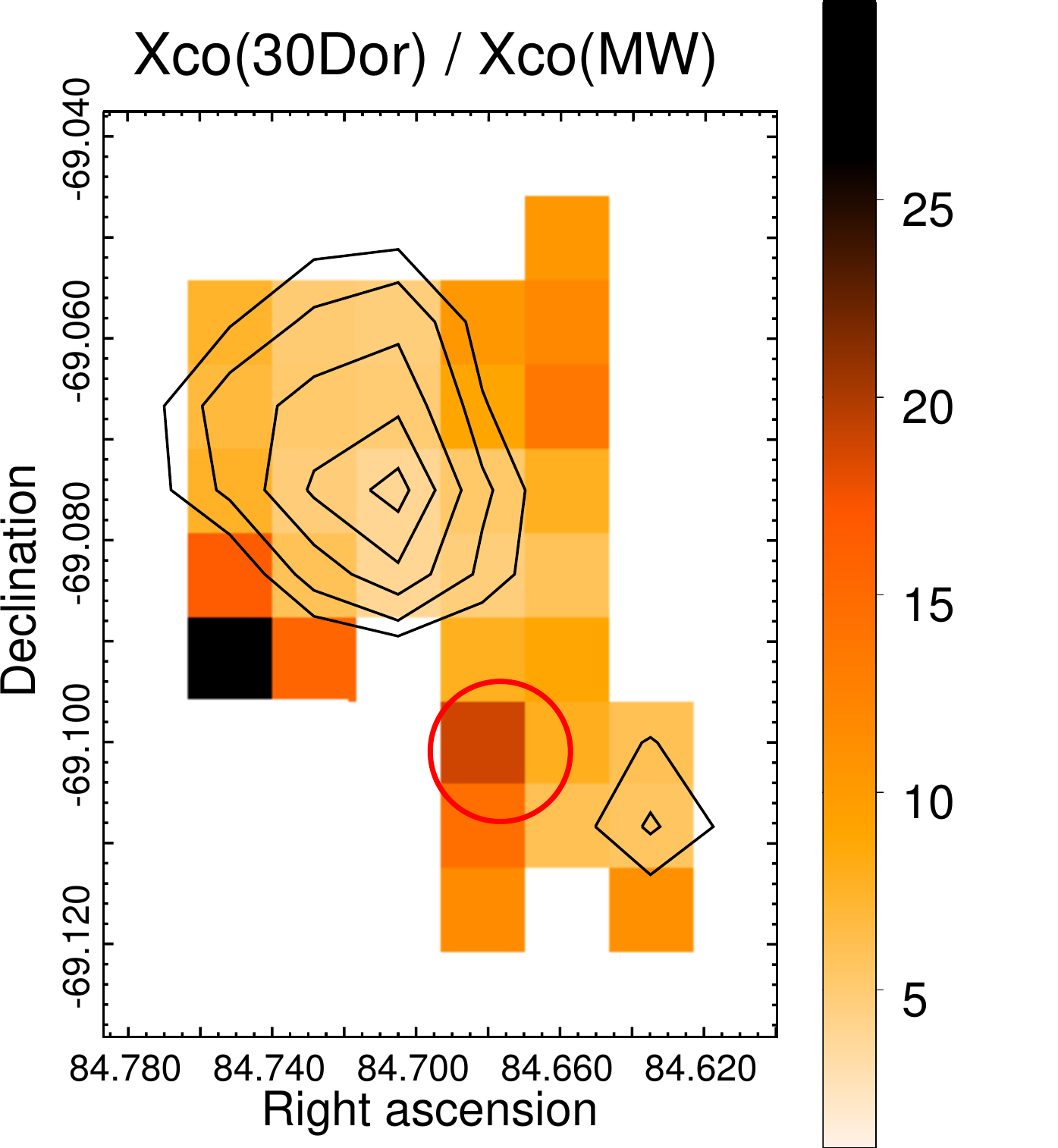}
     \caption{\textit{Left:} Mass surface density of \HH\ (in M$_{\odot}$ pc$^{-2}$) predicted by the PDR model. The black contours represent the $^{12}$CO (1-0) emission.
     The red circle indicate the location of the SSC R136. The maximum of the \HH\ mass surface density corresponds to the peak of CO intensity, as well as of \CII\ and \OI .
     \textit{Centre:} Fraction of the CO-dark gas calculated by comparing the predicted \HH\ column density from the Meudon PDR model and the estimation through the CO(1-0) observations and a galactic \Xcomw\ factor. The size of the circles indicates the total \HH\ mass surface density, between 30 and 220 M$_{\odot}$ pc$^{-2}$.
     \textit{Right:} Ratio of the conversion factors \Xcodor\ over \Xcomw .  The minima of \Xcodor\ correspond to the peaks of $^{12}$CO (1-0) (black contours). \Xcodor\ increases close to the location of R136 (red circle) and in the North-West region, where \CII\ is bright and were $P$ increases as well.}
              \label{fig:H2}
    \end{figure*}

From the Meudon PDR model, using the constrained parameters $P$, \go , \Avmax , and \ff\ from Section~\ref{sec:pdr_results} as inputs, we determine the column density and mass surface density of molecular gas predicted by the model for each 30\arcsec$\times$30\arcsec\ pixel. We estimate the \HH\ column density between 1.9$\times10^{21}$ and 1.3$\times10^{22}$ cm$^{-2}$ and the \HH\ mass surface density to be approximately between 30 and 220 M$_{\odot}$ pc$^{-2}$. The spatial distribution of the \HH\ mass surface density is presented in the left-hand panel of Figure~\ref{fig:H2}. The position of the maximum corresponds to the location of the \CII\ peak. It is also co-spatial with the CO (2-1) detected with ALMA by \cite{Indebetouw2013}. When integrating over the observed field of view, we find a total mass of \HH\ of $1.8\times 10^5 M_{\odot}$.

We compare the mass surface density of \HH\ estimated using the PDR modeling with the mass surface density of \HH\ that we can infer from the CO (1-0) emission at 43\arcsec\ resolution, using a standard galactic \Acomw\ conversion factor of 4.3 \Aunit\ (corresponding to \Xcomw\ $=2\times 10^{20}$ \Xunit ), which can be approximately applied down to metallicities of $\sim0.5$ Z$_{\odot}$ according to \cite{Bolatto2013}. Given the large uncertainties and the large scatter of the \Aco\ measurements, it is indeed not clear whether \Aco\ always scales with metallicity or if there is a plateau above $\sim 0.5 \Zsun$. The difference between the modeled \HH\ mass surface density from the Meudon PDR model and the CO-bright \HH\ mass surface density calculated from the CO (1-0) observations, normalised by the modeled total \HH\ mass surface density, is presented in the middle panel of Figure~\ref{fig:H2}. This quantity can be interpreted as the fraction of molecular gas not probed by CO (1-0), i.e.\ as the fraction of "CO-dark" molecular gas, and its spatial distribution will be discussed in more details in Section~\ref{sec:variation}. As shown in the middle panel of Figure~\ref{fig:H2} we find that 75-97\% of the total \HH\ mass surface density is associated with CO-dark gas within several tens of parsecs around the SSC R136. A large fraction of the molecular gas mass surface density is hence not probed when using CO (1-0) observations and a standard Milky Way \Xco\ factor in this half-metallicity, highly irradiated environment.

We can then calculate the equivalent \Xco\ conversion factor adapted for this region using:
\begin{equation}
\text{\Xcodor} = \frac{N(\text{\HH})_{\rm model}}{I_{\text{CO(1-0)}}}
\end{equation}
where $N(H_2)_{\rm model}$ is the \HH\ column density from the PDR model in cm$^{-2}$ and $I_{\rm CO(1-0)}$ is the observed CO intensity in K km s$^{-1}$. The \Xcodor\ conversion factor thus calculated ranges between 8$\times10^{20}$ and 4.0$\times10^{22}$ \Xunit\ (between $\sim 4$ and $\sim 20$ times \Xcomw ; right-hand panel of Figure~\ref{fig:H2}). We compare this range of values with previous literature measurements in Section~\ref{sec:litterature}.

\section{Discussion}
\label{sec:discussion}

\subsection{Effect of the spatial resolution}
\label{sec:resolution}

At the distance of the Andromeda galaxy ($\sim 780$ kpc ; \citealt{Ribas2005}) the entire region of $\sim$ 44 pc $\times$ 65 pc mapped here (limited by the overlap between the \OI\ and CO(1-0) field of views) would fall approximately in only one spaxel of the FIFI-LS spectrometer. To test whether the same analysis presented in this paper based on unresolved observations would lead to similar conclusions, we integrate all quantities over the covered area of 30Dor and repeat our analysis. Indeed, we have shown in \citet{Chevance2016} that the physical conditions derived from integrating over a region are typically biased toward the conditions of the high-pressure, highly-irradiated sub-regions, which might also affect the measured total \HH\ reservoir in an unresolved case, where different environments are mixed within a beam. 

We do not see a significant effect of the resolution on the measured \Xco\ factor or fraction of CO-dark gas (between $\sim$ 10 and 50 pc scale resolution). The integrated IR tracers over the full map are best reproduced by a PDR model with \go = 1390, $P$ = 6.18$\times 10^5$\,K\,cm$^{-3}$ and \Avmax $\sim 3$\,mag \citep{Chevance2016}. The \HH\ mass surface density predicted by a model with these particular input conditions is 194 $M_{\odot}$ pc$^{-2}$. It would therefore represent a total \HH\ mass of $3.4\times 10^5 M_{\odot}$ over the same region as mapped in Figure~\ref{fig:H2}. This is less than a factor of two higher than the total value of $1.8\times 10^5 M_{\odot}$ for the sum of the \HH\ mass estimated with the PDR model at 43\arcsec\ from Figure~\ref{fig:H2}. In addition, while the mean \Xcodor\ factor is $\sim 1.9 \times 10^{21}$\Xunit , the \Xcodor\ from the integrated model would be  $\sim 2.3 \times 10^{21}$\Xunit . This weak dependence of our measurement with the resolution could be due to the fact that we are still probing a relatively limited range of environments, under the dominant influence of R136.

\subsection{Comparison with previous \Xco\ estimations}
\label{sec:litterature}

In this section, we present some of the different methods commonly used to estimate the total mass of molecular gas. We compare our determination of the \Xco\ factor in 30Dor from Section~\ref{sec:pdr_H2} to values obtained from these different methods in previous studies. The results are summarised in Table~\ref{tab:litterature}.

  \subsubsection{Direct measurement of \HH\ in the warm gas}
 \label{sec:spitzer}

Observations of the warm \HH\ have been carried out with the Infrared Spectrograph (IRS; \citealt{Houck2004}), targeting \HH\ mid-infrared (MIR) lines (\HH\ 0-0 S(1), S(2) and S(3) transitions) in 30Dor at low- and high-resolution \citep{Lebouteiller2008, Indebetouw2009}. However, the low-resolution observations mostly provide upper limits only and the high-resolution observations are only available for a limited number of pointings. These limitations make it challenging to determine the distribution of the total molecular gas mass using direct \HH\ observations from IRS.

More recently, \HH\ 1-0 S(1) has been observed in the entire 30Dor nebula at 1\arcsec\ resolution with the CTIO 4-meter Blanco Telescope \citep{Yeh2015}. These high-resolution observations of the warm molecular gas in 30Dor reveal that the \HH\ emission originates from the PDR, and no evidence for shock excitation is found. The \HH\ mass inferred from these observations is obviously a strong lower limit on the total molecular gas mass as it does not include the mass associated with other levels of \HH\ and does not trace the reservoir of cold molecular gas associated with star formation.

 \subsubsection{\HH\ gas mass determination by dust}
 \label{sec:dustmodel}
 
Since direct \HH\ observations are challenging, the FIR emission from dust remains a commonly used tracer of the gas mass, through fitting of the dust spectral energy distribution (SED; e.g.  \citealt{Thronson1988, Israel1997, Dame2001, Leroy2007, Leroy2009, Leroy2011, Gratier2010, Bolatto2011}, in particular in dwarf galaxies). The gas-to-dust mass ratio (G/D) is believed to be fairly constant within a region and can be calculated as the ratio of the neutral gas mass ($M_{\text{\HI}}$) to the dust mass ($M_{\rm dust}$) at locations where neutral hydrogen dominates. 
Alternatively, the Galactic G/D value of $\sim 158$ \citep{Zubko2004} can be scaled with metallicity following \cite{Remy-Ruyer2014}. The \HH\ mass is then calculated as:
\begin{equation}
M_{\text{\HH}} = M_{\rm dust} \times (G/D) - M_{\text{\HI}}
\label{eq:Mdust}
\end{equation}
in an atomic or molecular gas dominated medium. This method has the advantage of being independent of CO emission observations, but the determination of the total gas mass from the dust mass is susceptible to variations of G/D. Especially, on small scales, this method relies on the assumption that gas and dust are well mixed. Even if this assumption is verified, the determination of the total gas mass still depends on the dust model used \citep[e.g.][]{Galliano2011, Remy-Ruyer2014}. In addition, while G/D roughly scales with metallicity \citep[e.g.][]{Sandstrom2013}, \cite{Remy-Ruyer2014} show that there is a large scatter in the G/D-metallicity relation between galaxies and therefore the scaling of G/D with metallicity is uncertain.

\begin{figure}
     \centering
      \includegraphics[trim=12cm 0mm 14mm 0mm, clip, width=8cm]{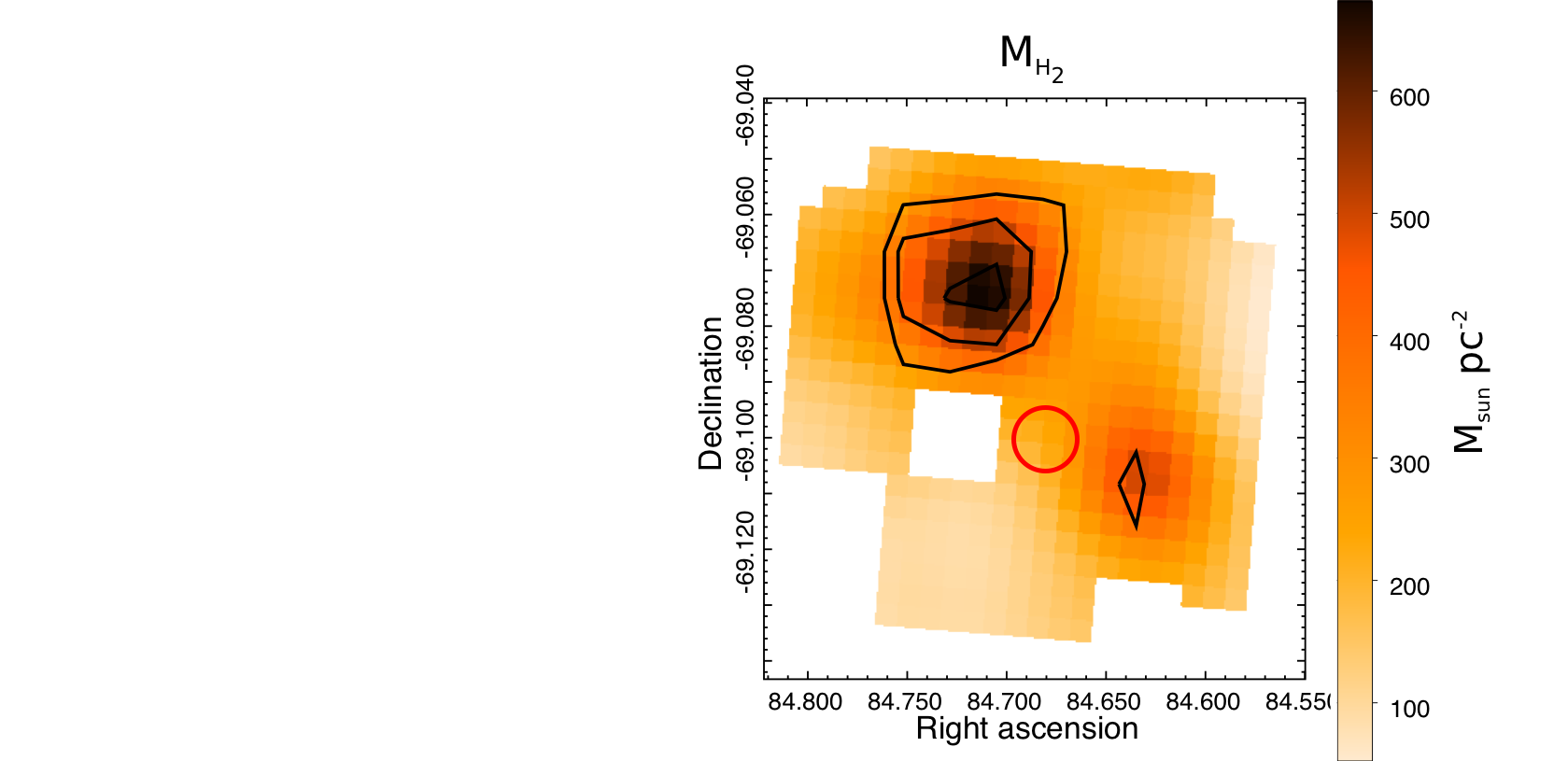}
     \caption{Mass surface density of molecular gas from the dust modeling. In black are the contours of the molecular mass surface density from the PDR modeling at 100-150-200 M$_{\odot}$ pc$^{-2}$. The red circle indicates the location of R136.}
     \label{fig:H2dust}
    \end{figure}

We apply this method to the 30Dor observations in order to compare the gas mass surface density determined from dust modeling with the results of Section~\ref{sec:pdr_H2} obtained with our PDR modeling.
We convolve the \spit /IRAC (3.6, 4.5, 5.8 and 8.0\mic\ bands) and MIPS (24 and 70\mic\ bands) and the \her /PACS (100 and 160\mic\ bands) and SPIRE (250\mic\ band) photometry data with a Gaussian function in order to degrade the resolution of the observations to 18\arcsec\ (similar to the resolution of SPIRE at 250\mic ). We use the phenomenological SED fitting procedure presented in \cite[Appendix C]{Galliano2018} to estimate the amount of dust along the line of sight for each pixel. This method uses a least-square fitter with the AC grain mixture of \cite{Galliano2011} and the diffuse interstellar radiation field intensity distribution from \cite{Dale2001}. This model was designed to fit the \her\ broadband photometry of the LMC, and to remain consistent with the observed elemental abundances. The free parameters include the total infrared luminosity (\Ltir ) between 3 and 1000\mic, the minimum starlight intensity, the difference between the maximum and minimum starlight intensities, the starlight intensity distribution power-law index and the PAH-to-total dust mass ratio. The G/D for 30Dor is determined using the broken power-law relationship between G/D and the metallicity defined in \cite{Remy-Ruyer2014}. The abundance ratio in 30Dor is $\log(n({\rm O})/n({\rm H})) = -3.75$ , as measured by \cite{Pellegrini2011}. This is in the regime where G/D scales linearly with the metallicity (this relation becomes non-linear for Z$\lesssim$0.1; \citealt{Remy-Ruyer2014}) and we derive (G/D)$_{\rm 30Dor} = 450$. Using this value of G/D and the \HI\ observations by \cite{Kim2003} in equation~\ref{eq:Mdust}, we find a mass surface density of molecular gas on average about a factor of 3 higher than that measured with our PDR modeling, with similar spatial variations (see Figure~\ref{fig:H2dust}). This leads to a range of \Xcodor\ factor of $3-12\times10^{21}$. The higher values of molecular gas mass surface density and \Xcodor\ found here in comparison to Section~\ref{sec:pdr_H2} can possibly be explained by an effective lower G/D in highly irradiated regions compared to the relationship calibrated by \cite{Remy-Ruyer2014} (here by a factor of $\sim 3$), or by the fact that we have considered an atomic or molecular gas dominated medium and we have not removed the possible contribution from the ionised gas. In addition, the above equation~\ref{eq:Mdust} only takes into account the scaling of G/D with metallicity, but not with gas density, which results from the accretion of material onto grains. This accretion in the dense ISM leads to an increase of the dust mass (up to a factor of 2), but also an increase of the emissivity of the grains by up to a factor of $\sim$ 3 \citep{Koehler2015}. This could therefore potentially lead to an overestimate of the dust mass by up to a factor of 6. Despite these caveats, the determination of the total \HH\ mass surface density from dust modeling also strongly suggest the presence of a large reservoir of CO-dark molecular gas.

  \begin{table*}
       \centering
       \caption{Measured CO-to-\HH\ conversion factor in the LMC and in star-forming (SF) regions of the LMC and SMC.}
       \begin{tabular}{llccc}
	  \hline
Reference                           & Method                    & \Xco\ [\Xunit]                    & Object                     & Physical size  \\ 
	 \hline 
   & PDR modeling  &   $0.8-4\times10^{21}$  & 30Dor  & 10\,pc\\
This study   & PDR modeling       &   $2\times10^{21}$      &  30Dor   & $\sim 50$\,pc \\
 & Dust modeling & $3-12\times10^{21}$ & 30Dor & 15\,pc\\[2.5ex]
\cite{Indebetouw2013}   & Virial equilibrium      &   $4\times10^{20}$    &  CO clumps in 30Dor              & 1\,pc\\[2.5ex]
\cite{Fukui2008}   & Virial equilibrium      &   $7\times10^{20}$    &  CO clouds in LMC              & 40\,pc\\[2.5ex]
   &   &   $8.4\times10^{21}$   & 30Dor   & 235\,pc\\ 
\cite{Israel1997}                 & Dust modeling  &   $1.3\times10^{21}$   & LMC SF regions                    & 235\,pc\\ 
   &  &   $12\times10^{21}$   & SMC SF regions  & 270\,pc\\[2.5ex]
\cite{Galametz2013}   &  Dust modeling      &  $5.4\times10^{20}$  & N159  &  10\,pc \\[2.5ex]
\cite{Roman-Duval2014}   & Dust modeling      &   <$6\times10^{20}$    &  LMC              & 15\,pc\\ [2.5ex]
\cite{Schruba2012}           & Constant depletion time         &   $4.6\times10^{21}$   & LMC                        & -- \\[2.5ex]
\cite{Bolatto2013}              &     Compilation                  &   $2\times10^{20}$      & LMC, Milky Way   & -- \\
	   \hline
        \end{tabular}
      \label{tab:litterature}
  \end{table*}
 
  \subsubsection{Other literature measurements on local scales}

Another classical method to estimate the mass of molecular gas is to apply the virial theorem to resolved CO clouds. This method is limited to the nearest galaxies due to resolution and sensitivity requirements. It has been in particular applied in the LMC \citep{Israel2003, Fukui2008, Hughes2010, Wong2011, Pineda2009}, leading to \Xco $\sim 8 \times 10^{20}$ \Xunit , relatively similar to the dust-based estimate from \citet[see Section~\ref{sec:globalcomparison}]{Roman-Duval2014}. However, this method likely leads to underestimation of the total \HH\ mass, and even more so in lower metallicity environments, as the CO core is likely to be smaller than the entire \HH\ cloud, due to the presence of a larger PDR envelope. As a result, virial measurements only consider the gas bright in CO and will not capture the potential CO-dark molecular gas reservoir. Besides, this method requires the CO molecular clouds to be resolved, and makes the assumption that the CO cores are in virial equilibrium, which is still debated \citep[e.g.][]{Ballesteros-Paredes2006}.

On smaller scales in 30Dor, \cite{Indebetouw2013} observed a $\sim$ 1\arcmin\ $\times$ 1\arcmin\ region in $^{12}$CO(2-1) and $^{13}$CO(2-1) with ALMA, centred on $\alpha = 5^{\rm h}38^{\rm m}48^{\rm s}$ and $\delta = -69^{\circ}0.4^{\prime}50^{\prime\prime}$ and also covered by our PACS and FIFI-LS observations. They estimate the molecular mass from the $^{12}$CO and $^{13}$CO intensities and calculate an \Aco\ factor of 8.9 \Aunit\ (corresponding to \Xco $\sim 4 \times 10^{20}$\Xunit ) for the mapped region. However, we note that these observations have high spatial resolution ($\sim 1$\,pc), and \cite{Indebetouw2013} measure specifically the \HH\ mass associated with the CO clumps. The \Xco\ factor including \HH\ possibly existing between the CO clumps is therefore expected to be higher. 

   \subsubsection{Comparison with literature measurements on global scales}
   \label{sec:globalcomparison}

On larger scales, \cite{Roman-Duval2014} use a single temperature blackbody modified by a broken-law emissivity to calculate the dust mass from the SED over the entire LMC, at a resolution of 15\,pc. They find an upper limit of $X_{\rm CO}^{\rm max} = 6 \times 10^{20}$ \Xunit\ for the LMC. Our values are about an order of magnitude higher than this upper limit, which can easily be explained by the fact that we are specifically focusing on a few tens of parsecs around R136, where the radiation field is the strongest and where the differential dissociation between CO and \HH\ is expected to be the most noticeable. In addition, the particular parameterisation of the dust model of \cite{Gordon2014} used by \cite{Roman-Duval2014} effectively results in an emissivity in the LMC significantly lower than in the Milky Way (i.e. a lower submillimeter excess in the LMC than in the MW), which leads to lower dust masses than measured here. At lower spatial resolution, \cite{Israel1997} presents dust-based estimates of the total molecular gas mass in various star-forming regions of the LMC and Small Magellanic Cloud (SMC). Their measured \Xco\ for the entire 30Dor region (over 235\,pc) is $8.4\times 10^{21}$\Xunit . This is again much higher than what would be expected from a simple scaling of the fiducial Milky Way value by the half-solar metallicity of the LMC and is consistent with our dust-based measurement, and with the upper range of our PDR-based measurement. The strong variation between the average \Xco\ factor found for the LMC and \Xcodor\ measured here and in \cite{Israel1997} indicates that it is not only the half solar metallicity ISM but the combination with the strong radiation from the SSC that together drives the large fraction of CO-dark molecular gas in our observations.

Another way to constrain the global \HH\ mass is by assuming a constant molecular gas depletion time. This method is based on the observed constant scaling between \HH\ surface density and the SFR surface density on large scales for nearby spiral galaxies.  \cite{Schruba2012} estimate the CO-to-\HH\ conversion factor in several dwarf galaxies of the IRAM HERACLES survey \citep{Leroy2009} making the assumption of a constant depletion time. They find \Aco\ $\sim 100$ \Aunit\  (i.e. \Xco\ $\sim 4.6 \times 10^{21}$ \Xunit ) for the integrated LMC. However, similarly to \Xco\ and G/D, the depletion time might vary with the environment. In addition, this method can only be applied on large scales, where observations average over a sufficiently large number of individual star-forming regions. On scales smaller than $\sim$ 1\,kpc, the correlation between \HH\ surface density and the SFR surface density breaks down, due to the insufficient statistical sampling of regions following independent evolutionary cycles \citep[e.g.][]{Schruba2010, KL14}. Specifically, the ratio between the \HH\ mass surface density and the SFR surface density of a single star-forming region (i.e. its instantaneous depletion time) varies a priori between 0, before the onset of star formation, and infinity, after the dispersion of the gas by feedback. This approach can therefore not be applied to determine the \HH\ mass in an individual star-forming region.

\subsection{Environmental variations of the CO-dark gas mass}
\label{sec:variation} 

In this section, we look for environmental factors that could explain the spatial variation of the CO-dark gas throughout 30Dor. We first calculate the average distance of the PDR in each pixel to the centre of R136, by comparing the emitted radiation field from the SSC, \gstar , to the incident radiation field on the PDRs, \go , which we derive from the PDR modeling as described in Section~\ref{sec:pdr_results}. We use for this the same approach as in \citet{Chevance2016}. We use the brightness temperature and position in the plane of the massive stars in R136 from the literature, and assume that they are randomly distributed along the line of sight, within a sphere of 6\,pc radius. We then propagate the ionising flux from each star, using a 1/$R^2$ relation (with $R$ the physical distance from the centre of the cluster), to calculate the average \gstar\ in a cube centred on R136, assuming no extinction along the line of sight. For each pixel of the map (i.e. each observed line of sight), we then estimate the distance $z$ of each pixel to the plane of R136 by matching \go\ to one of the possible values of \gstar\ along this line of sight. Finally, we calculate the 3D distance of each pixel to the centre of R136 as $R = \sqrt{z^2 + d^2}$, where $d$ is the projected distance between a given pixel and the centre of the source. 

\begin{figure*}
     \centering
     \includegraphics[trim=0mm 0mm 0mm 0mm, clip, width=8cm]{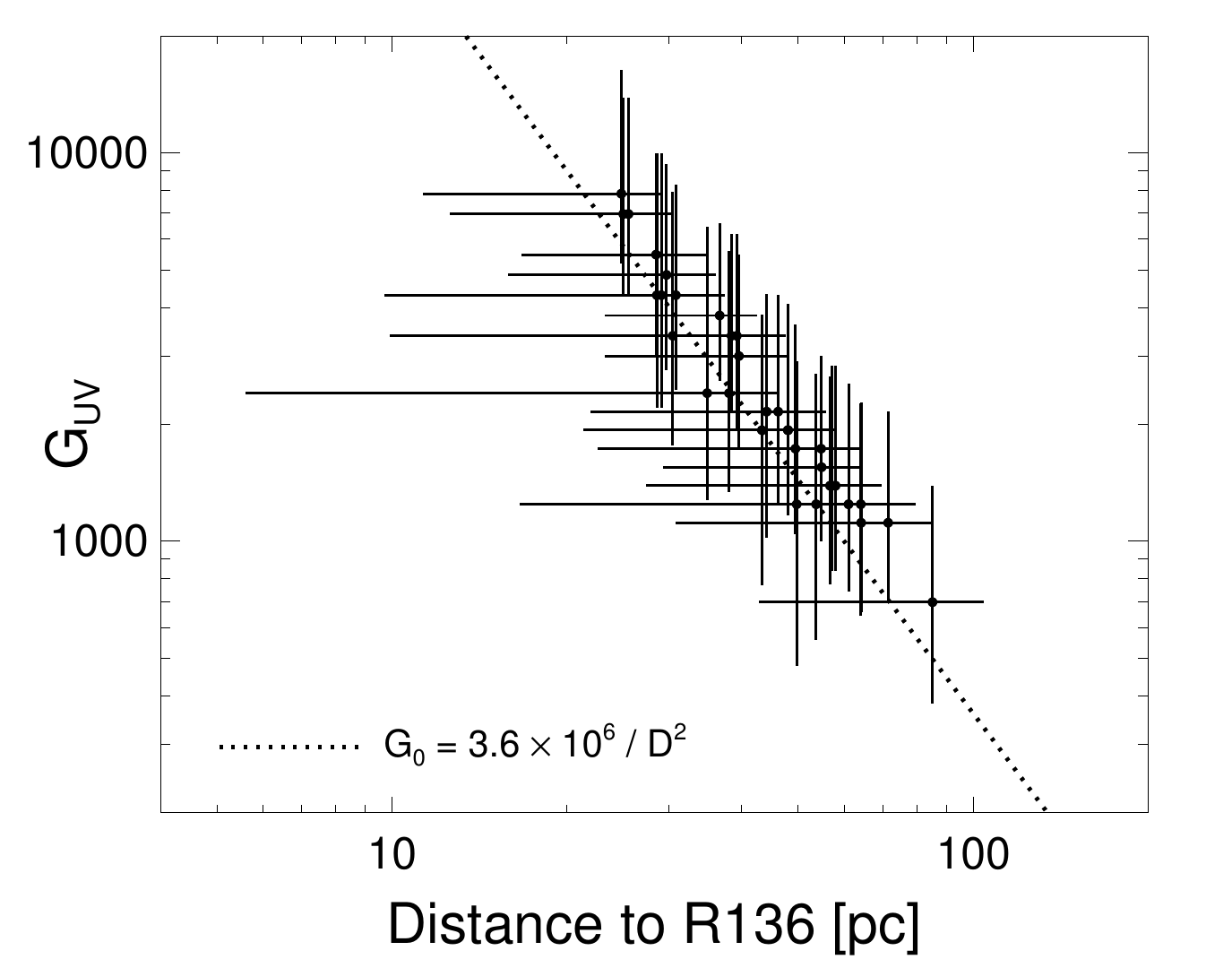}
     \includegraphics[trim=0mm 0mm 0mm 0mm, clip, width=8cm]{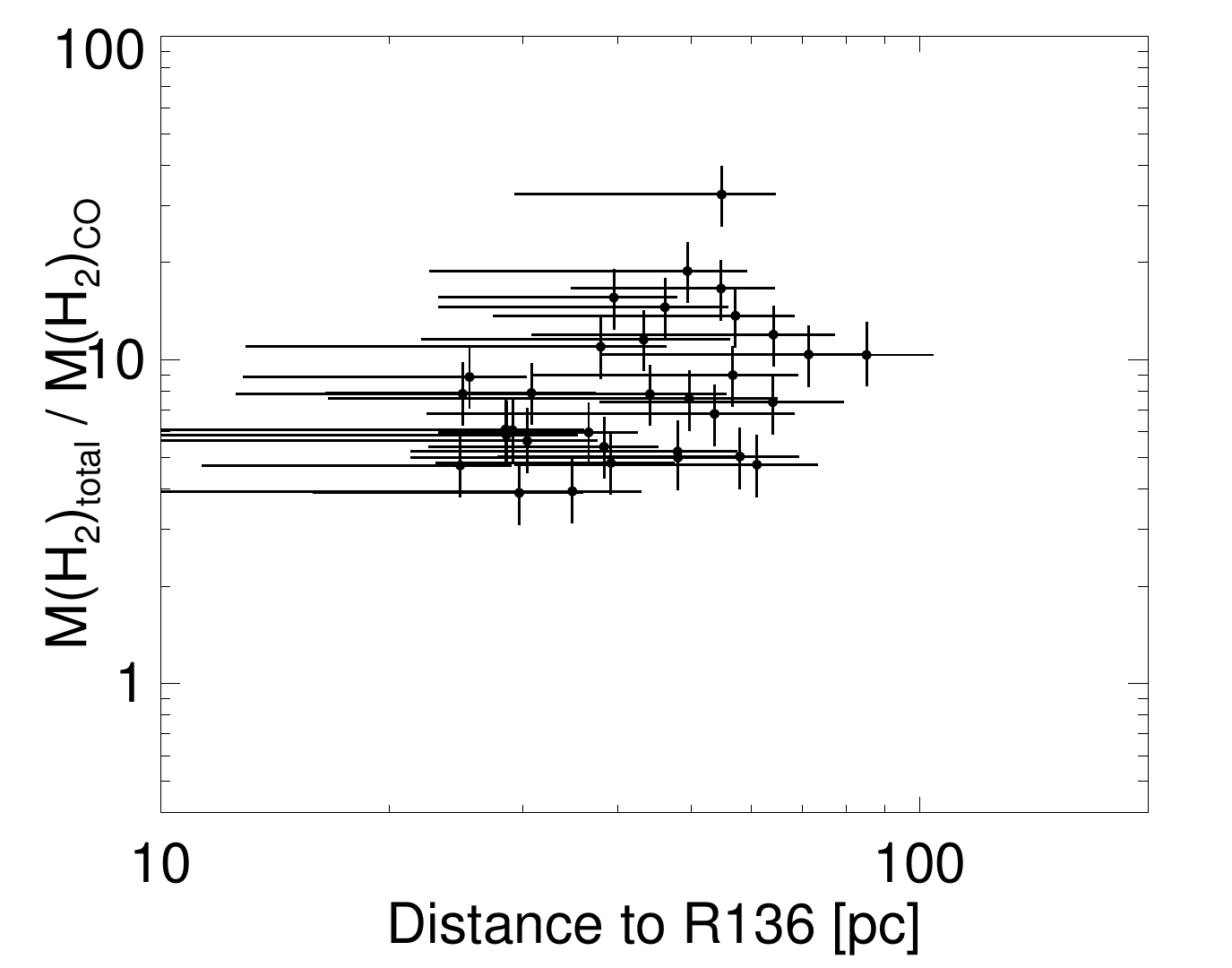}
     \includegraphics[trim=0mm 0mm 0mm 0mm, clip, width=8cm]{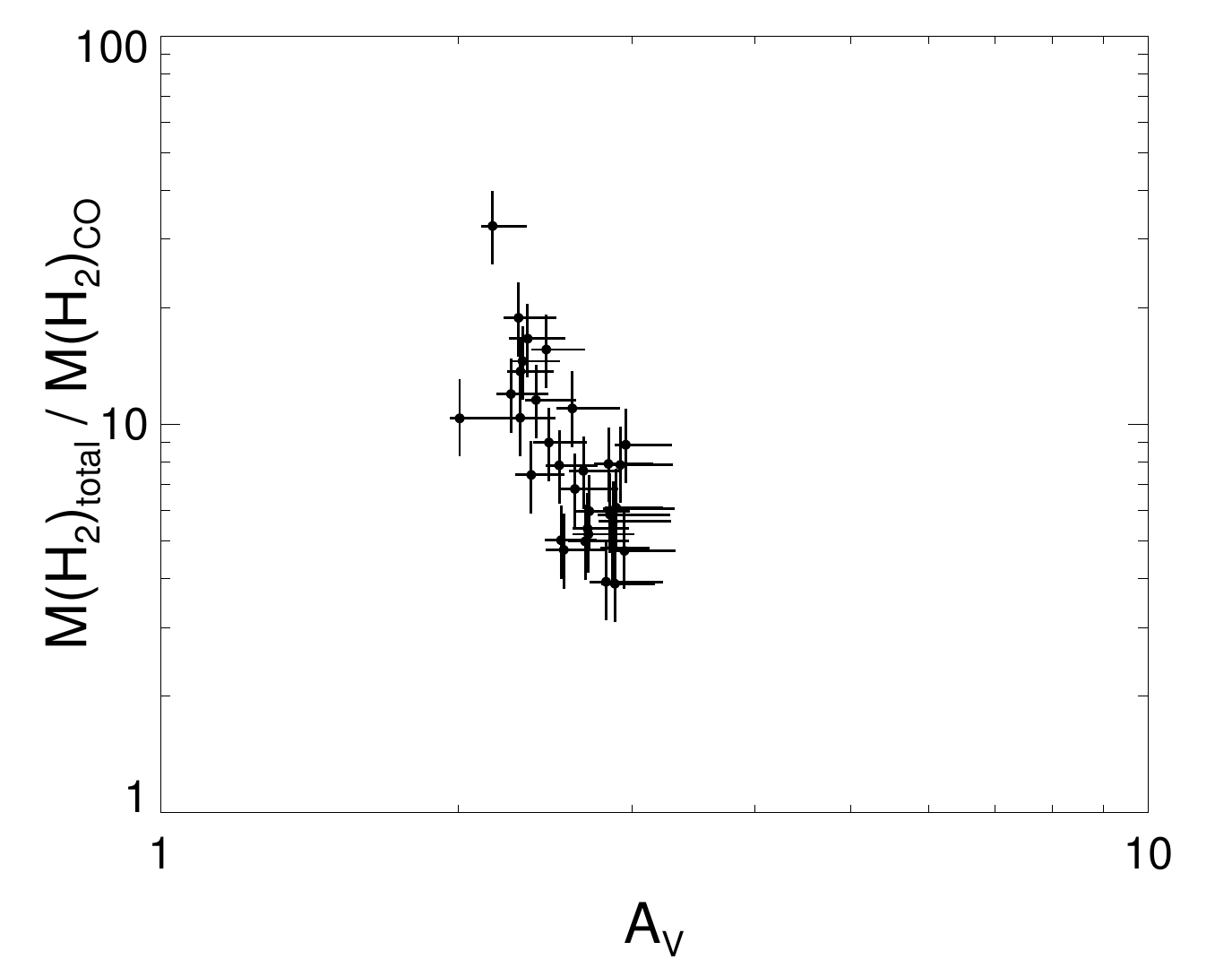}
      \includegraphics[trim=0mm 0mm 0mm 0mm, clip, width=8cm]{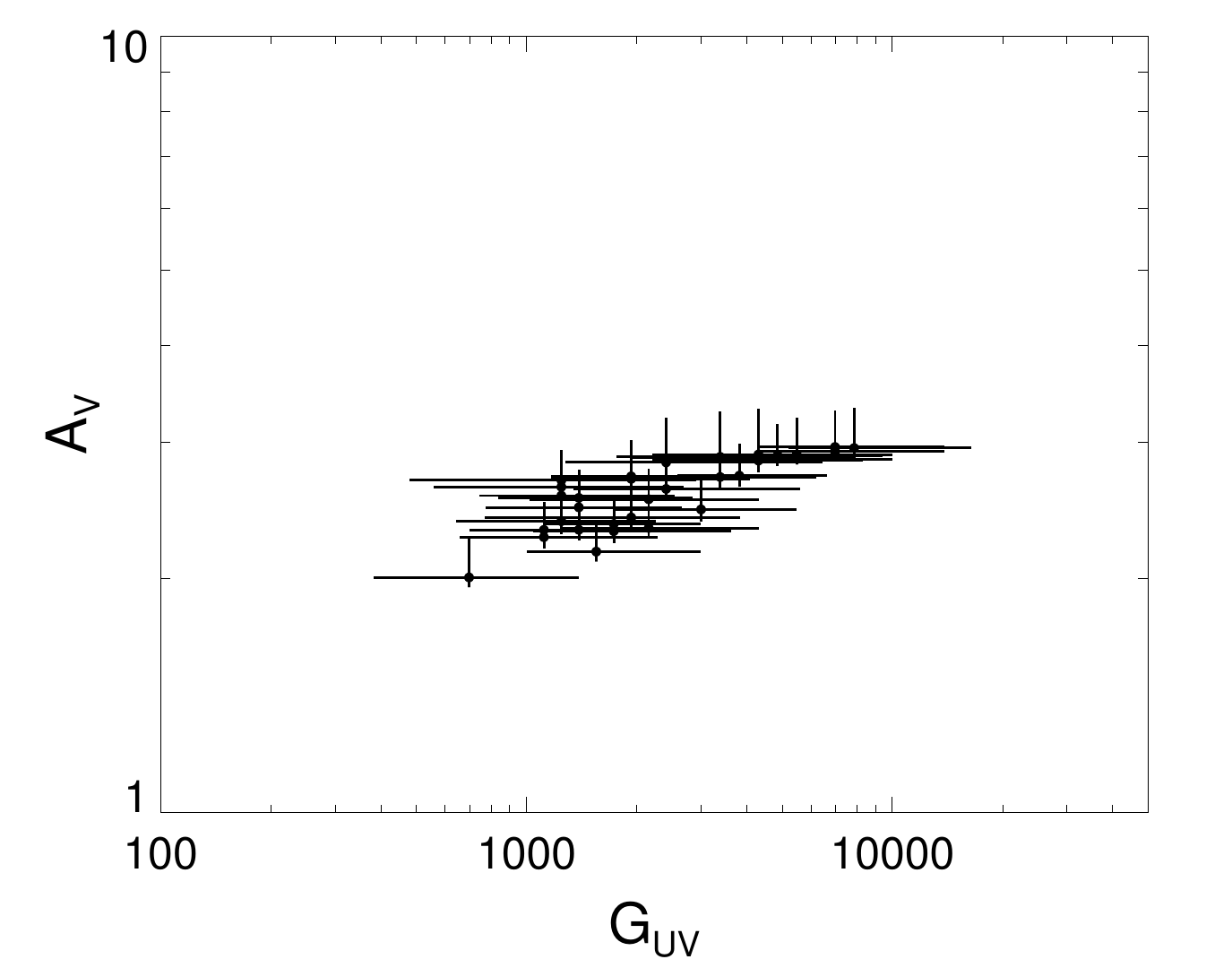}
     \caption{\textit{Top left:} \go\ as a function of the physical (deprojected) distance of the PDR clouds to the centre of R136. Clouds closer to the centre of the SSC receive a stronger radiation field. {The dotted line represents a 1/$D^2$ relation for the intensity of a central source with distance $D$, scaled to the mean emitted flux 30\,pc away form R136 (see text and \citealt{Chevance2016}). The fact that the data points lie slightly above this relation come from the spatial extension of the source, taken into account to infer the distances.}
     \textit{Top right:} CO-dark mass fraction as a function of the physical distance of the PDR clouds to the centre of R136. No significant correlation is observed.
     \textit{Bottom left:} CO-dark mass fraction as a function of the extinction in individual clouds. The observed anti-correlation is expected from the PDR model, as for clouds of smaller depth the (CO-dark) \HH\ layer in the PDR represents a larger fraction of the cloud. 
     \textit{Bottom right:} Extinction in individual clouds as a function of the incident radiation field. The observed correlation between \Av\ and \go\ compensates the expected increase of the CO-dark gas fraction with \go .}
              \label{fig:correlations}
\end{figure*}

One might expect an increase of the CO-dark gas fraction with the incident radiation field \go\ (or equivalently a decrease with the physical distance to R136). However, such a correlation is not observed in our results. While this may seem surprising, it can be explained by the observed anti-correlation between the CO-dark gas fraction and \Avmax , in combination with the observed decrease of \Avmax\ with distance to R136 (respectively decreasing \go ; see Figure~\ref{fig:correlations}), such that clouds with smaller shielding columns are observed towards the outskirts of the region. The first of these correlations is indeed expected from the model: the deeper the cloud, the better CO will trace the total \HH\ reservoir. On the contrary, for a small size cloud with a low \Av , the PDR layer including \HH\ but no CO represents a non negligible part of the cloud. In this context, the fact that we observe a decrease of \Avmax\ with increasing distance to R136 (respectively decreasing \go ) likely balances the expected increase of the CO-dark gas fraction with the intensity of the radiation field.

Through the strong correlation with \Avmax\ as noted above, there is a direct correlation between the CO-dark gas fraction and the ratio \CII /CO(1-0). This correlation is investigated further in Madden et al.\ (in prep.) by quantifying the total molecular gas mass on global scales in the Dwarf Galaxy Survey sources.

\section{Conclusions}
\label{sec:concl}

In this paper, we have presented new observations of the FIR lines \OIII\ 52\mic\ and 88\mic , \OI\ 145\mic\ and \CII\ 158\mic\ in the massive star forming region 30 Doradus of the LMC, observed during Cycles 4 and 5 of \sofia\ with FIFI-LS. These observations extend and complement previous \her /PACS observations of this region and allow us to explore the influence of the distance to the SSC R136 on the physical properties of the gas.

The spatially extended \OIII\ emission and the overall low values of the ratio \CII /\OIII\ 88\mic\ reveal a porous medium where high-energy photons emitted by R136 propagate over several tens of pc before interacting with the ISM. The ratio \OIII ~52\mic /\OIII ~88\mic , previously inaccessible with \her , ranges between 0.48 and 1.84 in 30Dor, and is a good probe of the electron density in the low-density ionised gas. Overall, we measure a maximum density of $\sim 600$\,cm$^{-3}$ around R136. This is in agreement with previous measurements of the electron density of the ionised gas in this region \citep[e.g.][]{Indebetouw2009, Pellegrini2011, Chevance2016}, although the different line ratios used in these studies might probe different phases of the ionised gas. 

Following a similar approach as in \citet{Chevance2016}, we have determined the spatial distribution of the pressure and radiation field in 30Dor, by fitting our observations with the Meudon PDR model. We find that the incident radiation field \go\ ranges between $7\times10^2$ and $7\times10^3$ (in units of the standard radiation field defined in \citealt{Mathis1983}) and the pressure $P$ ranges between $4.3\times10^5$ and $1.6\times10^6$\,K\,cm$^{-3}$, at a resolution of 43\arcsec\ ($\sim$ 10\,pc). By making use of the ratio \CII /CO(1-0), which is sensitive to the depth of the PDR clouds, we determine a maximum total visual extinction \Avmax\ of 2 to 3\,mag.

In 30Dor, the intense radiation field induced by the massive excitation source R136 on the surrounding half-solar metallicity gas creates an extreme environment. Based on the physical properties of the gas determined through the PDR modeling, and by comparing to CO(1-0) observations, we estimate that between 75\% and 97\% of the \HH\ would be undetected, and therefore can be considered as ``CO-dark'' molecular gas, if a standard \Xcomw\ factor of $2\times10^{20}$\Xunit\ would be applied in this region. As a result, a more appropriate CO-to-\HH\ conversion factor around R136 is \Xcodor $\sim 1-4 \times 10^{21}$ \Xunit , in agreement with previous determinations of this conversion factor in 30Dor and other regions of the LMC using different methods. We note that ``CO-dark'' gas here does not presume whether the CO lines are simply too faint relatively to the Milky Way environment, or if a significant fraction of the \HH\ gas is CO-free and co-spatial with ionised and neutral carbon. This question is investigated in more detail in Madden et al.\ (in prep.).

The spatial variation of the CO-dark fraction throughout the mapped regions is mostly driven by the amount of visual extinction \Avmax\ (i.e.\ the depth of the clouds). This strong dependence seems to balance any expected trend with the incident radiation field or the distance to R136. The influence of the SSC on the neutral and molecular gas is therefore still dominant out to several tens of parsecs away from the centre of the cluster. Future observations of other star-forming (or more diffuse) regions in the LMC and SMC would allow us to disentangle the combined effects of the radiation field and the metallicity on the properties of the ISM (e.g. the pressure, porosity, and size of the PDR clouds) and ultimately probe the environmental dependence of the CO-dark gas fraction.

\section*{Acknowledgements}
We thank the referee, Neal Evans, for constructive feedback that improved this paper.
The authors would like to thank Annie Hughes for providing the CO(1-0) data.
MC and JMDK gratefully acknowledge funding from the Deutsche Forschungsgemeinschaft (DFG, German Research Foundation) through an Emmy Noether Research Group (grant number KR4801/1-1) and the DFG Sachbeihilfe (grant number KR4801/2-1).
JMDK gratefully acknowledges funding from the European Research Council (ERC) under the European Union's Horizon 2020 research and innovation programme via the ERC Starting Grant MUSTANG (grant agreement number 714907).
DC is supported by the European Union's Horizon 2020 research and innovation programme under the Marie Sk\l{}odowska-Curie grant agreement No 702622. 
FLP is supported by the ANR grant LYRICS (ANR-16-CE31-0011). 
The authors would like to thank the French Programme National "Physique et Chimie du Milieu Interstellaire" (PCMI) of CNRS/INSU with INC/INP co-funded by CEA and CNES, for supporting our SOFIA flights and collaboration with SOFIA projects.
This research was made possible through the financial support of the Agence Nationale de la Recherche (ANR) through the programme SYMPATICO (Program Blanc Projet NR-11-BS56-0023) and through the EU FP7.
This work is based on observations made with the NASA/DLR Stratospheric Observatory for Infrared Astronomy (SOFIA). SOFIA is jointly operated by the Universities Space Research Association, Inc. (USRA), under NASA contract NNA17BF53C, and the Deutsches SOFIA Institut (DSI) under DLR contract 50 OK 0901 to the University of Stuttgart.

\bibliographystyle{mnras}
\bibliography{biblio}


\bsp
\label{lastpage}
\end{document}